\documentclass[sigconf]{acmart}
\AtBeginDocument{%
  }

\setcopyright{acmlicensed}
\copyrightyear{2018}
\acmYear{2018}
\acmDOI{XXXXXXX.XXXXXXX}
\acmConference[Conference acronym 'XX]{Make sure to enter the correct
  conference title from your rights confirmation email}{June 03--05,
  2018}{Woodstock, NY}
\acmISBN{978-1-4503-XXXX-X/2018/06}

\newcommand{\blocktitle}[1]{\noindent\textbf{#1}\\[0.4em]} 
\newcommand{\blockgap}{}                     

\usepackage{enumitem}
\usepackage{placeins}

\setlist[itemize]{leftmargin=15pt}
\setlist[enumerate]{leftmargin=15pt}

\usepackage{float}   
\usepackage{tabularx}

\begin{document}

\title{Credibility Governance: A Social Mechanism for \\ Collective Self-Correction under Weak Truth Signals}

\author{Wanying He}
\affiliation{%
  \institution{School of Artificial Intelligence, Wuhan University; BIGAI}
  \city{Wuhan}
  \country{China}
  }
\email{wanying.he@whu.edu.cn}

\author{Yanxi Lin}
\affiliation{%
  \institution{Tsinghua University}
  \city{Beijing}
  \country{China}
}
\email{linyx22@mails.tsinghua.edu.cn}

\author{Ziheng Zhou}
\affiliation{%
  \institution{University of California, Los Angeles}
  \city{Los Angeles}
  \country{US}
  }
\email{josephziheng@ucla.edu}

\author{Xue Feng}
\affiliation{%
  \institution{State Key Laboratory of General Artificial Intelligence, BIGAI}
  \city{Beijing}
  \country{China}
  }
\email{feng.xue1580@gmail.com}

\author{Min Peng}
\affiliation{%
  \institution{School of Artificial Intelligence, Wuhan University}
  \city{Wuhan}
  \country{China}
  }
  \email{pengm@whu.edu.cn}

\author{Qianqian Xie}
\affiliation{%
  \institution{School of Artificial Intelligence, Wuhan University}
  \city{Wuhan}
  \country{China}
  }
\email{xqq.sincere@gmail.com}

\author{Zilong Zheng}
\affiliation{%
  \institution{State Key Laboratory of General Artificial Intelligence, BIGAI}
  \city{Beijing}
  \country{China}
  }
\email{zlzheng@bigai.ai}

\author{Yipeng Kang}
\affiliation{%
  \institution{State Key Laboratory of General Artificial Intelligence, BIGAI}
  \city{Beijing}
  \country{China}
  }
  \email{kangyipeng@bigai.ai}








\begin{abstract}
Online platforms increasingly rely on opinion aggregation to allocate real-world attention and resources, yet common signals such as engagement votes or capital-weighted commitments are easy to amplify and often track visibility rather than reliability. This makes collective judgments brittle under weak truth signals, noisy or delayed feedback, early popularity surges, and strategic manipulation. We propose \emph{Credibility Governance} (CG), a mechanism that reallocates influence by learning which agents and viewpoints consistently track evolving public evidence. CG maintains dynamic credibility scores for both agents and opinions, updates opinion influence via credibility-weighted endorsements, and updates agent credibility based on the long-run performance of the opinions they support, rewarding early and persistent alignment with emerging evidence while filtering short-lived noise. We evaluate CG in \textsc{POLIS}, a socio-physical simulation environment that models coupled belief dynamics and downstream feedback under uncertainty. Across settings with initial majority misalignment, observation noise and contamination, and misinformation shocks, CG outperforms vote-based, stake-weighted, and no-governance baselines, yielding faster recovery to the true state, reduced lock-in and path dependence, and improved robustness under adversarial pressure. Our implementation and experimental scripts are publicly available at \texttt{https://github.com/Wanying-He/Credibility\_Governance}.
\end{abstract}


  


\maketitle

\section{Introduction}

Imagine you are a young researcher exploring a cross-disciplinary idea that shows quiet promise. You release a short working paper with cautious but intriguing pilot results, and discussion begins in an web forum where your field increasingly exchanges ideas. The signals you can gather are slow, noisy, and hard to interpret, yet the direction feels meaningful. Over time, however, the conversation on the platform starts to drift. A different line of research, championed by a tightly connected group of scholars, produces a rapid stream of claims and polished updates. They highlight one another’s progress, repost each other’s summaries, and their coordinated activity shapes the platform’s opinion statistics that funding bodies and program committees increasingly monitor. Although the underlying evidence remains thin, the visibility and volume of support for this competing direction grow quickly, and aggregated signals on the platform begin to guide invitations, collaborations, and resource allocation. Your more careful line of inquiry, despite grounded in more reliable signals, becomes harder to detect within the aggregated consensus, and later scholars inherit a distorted sense of what the field believes, confounding their judgments.

\addvspace{0.3em}
This scenario reflects a systemic limitation of modern social and web based platforms. When attention is aggregated through rapid, reinforcing social signals, slower yet more reliable indications of quality struggle to surface \cite{muchnik2013, bakshy2015}. As a result, collective evaluations can drift towards what is most visible rather than what is best supported, and communities may converge prematurely on directions that later prove fragile \cite{lorenz2011}.

\addvspace{0.3em}
This paper asks whether a collective can steer itself away from such distortions when truth is only weakly observable. We propose a mechanism called \emph{Credibility Governance} (CG), designed for environments where agents must infer truth from uncertain, delayed, and uneven signals \cite{golub2010}. Rather than treating visibility or raw support as evidence, CG allocates influence and updates credibility dynamically according to how agents interpret changing conditions over time, aiming to preserve sensitivity to emerging evidence without amplifying transient fluctuations.

\addvspace{0.3em}
Our contributions are threefold. First, we introduce a dual-component simulation environment that links a physical world of topic progress with a social world where agents form, express, and aggregate opinions, allowing systematic study of truth seeking under partial observability. Second, we formalize Credibility Governance and compare it with three baseline aggregation rules: a staking-based mechanism that amplifies topics backed with greater resource commitment, a vote-based mechanism that reflects simple majority preferences, and a no-governance baseline that applies collective choices directly to observed progress. Together, these baselines capture dominant logics through which contemporary platforms aggregate opinions \cite{sunstein2006}. Third, we show that CG supports collective self correction across a broad range of noisy, delayed, or misleading environments, while also identifying boundary conditions that clarify when the mechanism succeeds and where it requires refinement.

\addvspace{0.3em}
Taken together, these contributions offer a step towards opinion aggregation systems that remain epistemically stable even when evidence is sparse, delayed, or difficult to verify \cite{kleinberg2022trust}.

\begin{figure}
  \centering
  \includegraphics[width=1\linewidth]{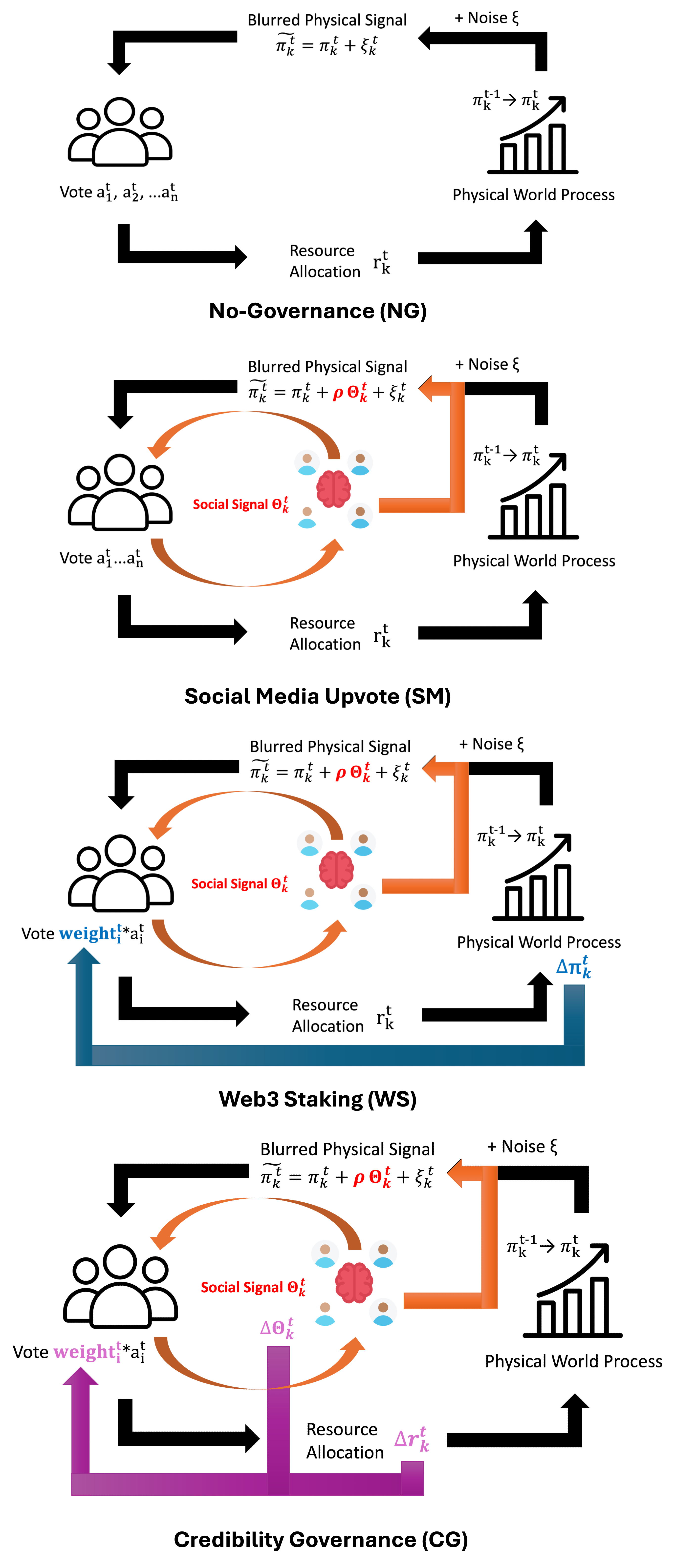}
  \caption{Comparison of the four governance mechanisms.\\
  (1) NG: Agents observe physical signals \(\tilde{\pi}_k^{t}\). \\
  (2) SM: NG + social signal \(\Theta_k^{t}\) in observation.\\
  (3) WS: SM + agent influence $w_i^t$ updated by \(\Delta \pi_k^t\).\\
  (4) CG: WS + agent influence $w_i^t$ updated by \(\Theta_k^{t}\).}
  \label{fig:main}
\end{figure}

\section{Methodology}
    \subsection{The POLIS Simulation Framework}
    \addvspace{0.2em}
    \blocktitle{2.1.1 Platform Capabilities}
    To support rigorous evaluation of Credibility Governance, we introduce \textbf{POLIS}, a modular simulation platform designed for studying the co-evolution of social dynamics and real-world outcomes. POLIS coordinates large populations of LLM-driven agents interacting across a coupled \textbf{Physical World} and \textbf{Opinion World}, and provides the following core capabilities:

    \begin{enumerate}
        \item\textbf{{Scalable Multi-Agent Orchestration.}}
        The platform can manage thousands of autonomous LLM agents, each equipped with persistent identities, memory, and reasoning stability.
        \item\textbf{{Parametric Physical World Modeling.}}
        The Physical World is fully parametric and defines flexible resource-to-progress mappings that translate collective decisions on resource allocations into real-world outcome trajectories. This design supports arbitrary progress functions, enabling POLIS to emulate diverse domains such as scientific discovery, technological adoption, or policy implementation.
        \item\textbf{{Modular Governance Interface.}}
        POLIS supports pluggable governance modules that define how the Opinion World aggregates agent opinions into collective decisions and updates social states. This makes it easy to compare different governance mechanisms under same setups of agent and world .
        \item\textbf{{Comprehensive Data Logging.}}
        Every agent decision, rationale, state change, and environmental signal is logged, enabling fine-grained, reproducible analysis of both micro-level behaviors and macro-level emergent phenomena.
    \end{enumerate}

    \addvspace{1.0em}
    \blocktitle{2.1.2 Dual-World Structure and Interaction}
    The \emph{Physical World} and \emph{Opinion World} are linked through a shared set of \textbf{topics}. Agents form perceptions from public signals, vote on topics, and their collective choices determine subsequent topic progress. The following components outline the key elements of each world and sketch how information flows between them.
    
    \paragraph{\textbf{(1) Topics: The Core Cross-World Concept}}
    Topics represent competing real-world lines of effort, such as scientific hypotheses or tech pathways. Each topic maintains attributes in both worlds:
    \begin{itemize}
        \item \(\pi_k^t\): a \emph{physical state} represented by its cumulative progress value,
        \item \(\Theta_k^t\): an \emph{opinion-state representation} including its current support level and publicly visible social signal.
    \end{itemize}
    As agents vote \emph{on topics} and mechanisms act on their support, topics provide the essential interface connecting the two worlds.
    
    \paragraph{\textbf{(2) Physical World: Topic Progress Dynamics}}
    The Physical World governs how resource allocations translate into measurable progress. Each topic \(k\) has a cumulative progress value \(\pi_k^t\), and its increment \(\Delta \pi_k^t\) is determined by a parametric update rule that specifies the topic’s development trajectory. In this work, we allow topics to evolve through nonlinear trajectories that can transition across multiple growth phases, yielding varied patterns such as slow starts, rapid gains, or diminishing returns. Environmental fluctuations enter through additive noise terms, capturing real-world uncertainty in progress.
    
    \paragraph{\textbf{(3) Opinion World: Agent and Topic Social States}}
    The Opinion World maintains the evolving social layer of the system. Each agent tracks:
    \begin{itemize}
        \item $\alpha_i^t$: beliefs and confidence on topics derived from public signals,
        \item $a_i^t$: its current vote choice over topics,
        \item $w_i^t$ agent influence weights, derived from mechanism-specific attributes such as credibility or stake balances.
    \end{itemize}
    Topics maintain a social signal \(\Theta_k^t\), capturing both their current support level and the rate at which that support evolves, as defined by the chosen governance mechanism. 
    
    \paragraph{\textbf{(4) Physical to Opinion: Agent Perception}}
    Agents do not observe the true physical state directly. Instead, they receive a social signal \(\Theta_k^{t-1}\) and a noisy physical signal \(\tilde{\pi}_k^{t-1}\), designed to emulate real-world settings where ground truth is inaccessible and most observable evidence is distorted by social influence. These signals constitute the agent’s perceived evidence, shaping its worldview and informing its support for different topics.
        
    \paragraph{\textbf{(5) Opinion to Physical: Vote Aggregation}}
    Agents cast votes \(a_i^t\) \emph{on topics} based on their beliefs and epistemic personas. A governance mechanism then aggregates these votes into a resource allocation vector \(\mathbf{r}^t\) through mechanism-specific rules. This allocation is then converted into topic progress increments \(\Delta \pi_k^t\), completing the cross-world feedback cycle.

    \addvspace{1.0em}
    \blocktitle{2.1.3 Round-Level Walkthrough}
    A simulation round describes how the Physical World and Opinion World evolve from time \(t-1\) to \(t\). The process is strictly causal: agents observe signals produced at the end of round \(t-1\), form perceptions, vote on topics, and their collective decisions update both physical progress and the opinion-state variables that will generate the signals observed in the next round. A walkthrough of the round structure is as follows:
    
    \paragraph{\textbf{(1) Perception:}}
        At the start of round \(t\), each agent observes the previous round’s social signal \(\Theta_k^{t-1}\) and the previous round’s noisy physical signal \(\tilde{\pi}_k^{\,t-1}\). Together, these signals constitute the agent’s perceived evidence and shape its worldview for round \(t\).
    
    \paragraph{\textbf{(2) Opinion Formation and Aggregation:}}
        Each agent updates its \(\alpha_i^t\), including its \emph{beliefs}, a qualitative choice of which topic to support, and its \emph{confidence}, a quantitative measure of how strongly it holds that view. This yields a voting intention \(a_i^t\) over a chosen topic. A governance mechanism then aggregates these intentions into a resource allocation vector \(\mathbf{r}^t\) using influence weights \(w_i^{\,t-1}\).
        
    \paragraph{\textbf{(3) State Update:}}
        Given \(\mathbf{r}^t\), both worlds update their states:
        \begin{itemize}[leftmargin=30pt]
            \item \textbf{Physical World} converts \(\mathbf{r}^t\) into topic progress increments \(\Delta \pi_k^t\), updating cumulative progress values \(\pi_k^t\). 
    
            \item \textbf{Opinion World.}  
            The governance mechanism updates agent influence weights \(w_i^t\) and topic-level social states, yielding the next-round social signal \(\Theta_k^{t}\).
            Using the updated progress and social signal, the system then constructs the next round’s noisy physical signal
            \[
                \tilde{\pi}_k^{\,t} = \pi_k^{\,t} + \rho\,\Theta_k^{\,t} + \xi_k^{\,t},
            \]
            which, together with \(\Theta_k^t\), will be observed by agents at the start of round \(t+1\).
        \end{itemize}

    \subsection{Opinion World (Governance Mechanisms)}
    To evaluate Credibility Governance, we compare it against three representative aggregation rules: (1) a Web3-style Staking (WS) mechanism that instantiates resource-weighted governance, (2) a Social Media Upvote (SM) mechanism that instantiates engagement-driven voting, and (3) a No-Governance baseline in which agents act solely on observed signals. 

    \addvspace{0.3em}
    All four mechanisms evaluated are instantiated from the same protocol detailed above (visualized in Figure~\ref{fig:main}), but differ in how they assign influence weights \(w_i^t\), how they update the social signal \(\Theta_k^t\), and what their influence update is based on (social momentum or physical progress). Key differences across mechanisms are summarized in Table~\ref{tab:mech_comparison}.
    
    \begin{table*}[t]
    \centering
    \caption{Comparison of governance mechanism properties.}
    \label{tab:mech_comparison}
    \renewcommand{\arraystretch}{1.08}
    \setlength{\tabcolsep}{6pt}
    \footnotesize
    \begin{tabularx}{\textwidth}{p{2.6cm} XXXX}
    \toprule
    \textbf{Feature} &
    \textbf{Credibility Gov. (CG)} &
    \textbf{Web3 Staking (WS)} &
    \textbf{Social Media (SM)} &
    \textbf{No Governance (NG)} \\
    \midrule
    Agent Influence (\(w_i\)) &
    Credibility-based &
    Stake-based &
    Equal influence &
    Equal influence \\

    Social Signal Update (\(\Theta_k\)) &
    Cred.-adj. \(\Delta\)support &
    Stake-wtd popularity (delay \(\tau\), noise \(\xi\)) &
    Raw popularity &
    None \\

    Signal distortion &
    noise \(\xi\), contam. \(\rho\Theta_k^t\) &
    noise \(\xi\), delay \(\tau\) (stake settles) &
    noise \(\xi\), contam. \(\rho\Theta_k^t\) &
    noise \(\xi\) \\

    Robustness knob &
    anti-bubble \(B_k^t\) &
    stake cost / budget &
    None &
    None \\

    Reward Basis &
    \(\Delta\Theta_k^t\) &
    \(\Delta\pi_k^t\) (delayed) &
    N/A &
    N/A \\
    \bottomrule
    \end{tabularx}
    \end{table*}
    
    \paragraph{Interpretation of the social signal \(\Theta\).}
    We use \(\Theta_k^t\) as an aggregate, publicly observable proxy for the crowd's current support for option or topic \(k\) at time \(t\) (e.g., the platform-level ``consensus'' signal). In many human systems, ground truth is revealed only weakly, late, or through noisy proxies, while \(\Theta_k^t\) is available immediately and shapes behavior. We therefore interpret \(\Theta_k^t\) as a \emph{public evidence proxy}, in the sense that it summarizes what information appears to be accumulating in the population at that moment. Crucially, CG does not reward high support per se; it rewards \(\Delta\Theta_k^t\), the \emph{change} in support over time. This ``filtered momentum'' choice emphasizes informative movement, such as belief revision after new evidence, rather than reinforcing static popularity levels. When truth signals are weak, \(\Delta\Theta_k^t\) provides a lightweight, behaviorally grounded proxy for evidence accumulation, allowing CG to increase influence for agents who consistently contribute to constructive shifts in public belief.

    \noindent\textbf{Failure modes.}
    This proxy can fail when social movements are dominated by manipulation or coordination rather than evidence, or when observation noise becomes so large that \(\Theta_k^t\) no longer tracks underlying outcomes, motivating our noise and adversarial stress tests in Section~\ref{sec:results}.


    \addvspace{1.0em}
    \blocktitle{2.2.1 Credibility Governance (CG)}
    CG assigns agent influence based on epistemic performance rather than raw support counts. Agents gain 
    credibility when the topics they supported in the previous round receive a higher social signal 
    \(\Theta_k^t\), and lose credibility when their earlier choices move against the emerging consensus. This 
    way, CG ties influence to an agent’s historical alignment with credible shifts in public evidence.
    
        \paragraph{\textbf{(1) Perception:}}
        Each agent observes the previous round’s social signal \(\Theta_k^{t-1}\) (calculation defined in State Update) and noisy physical signal \(\tilde{\pi}_k^{\,t-1}\), forming its perceived evidence for round \(t\).
    
        \paragraph{\textbf{(2) Opinion Formation and Aggregation:}}
        Agents update their beliefs (topic preference) and confidence \(\alpha_i^t\) (strength of preference), producing a voting intention \(a_i^t\).  
        Influence weights are credibility-weighted:
        \[
            w_i^{t-1} = \alpha_i^t \exp(\lambda c_i^{\,t-1}),
        \]
        and the allocation is:
        \[
            r_k^t = 
            \frac{\sum_i w_i^{t-1}\mathbf{1}[a_i^t = k]}
                 {\sum_j w_j^{t-1}}.
        \]
    
        \paragraph{\textbf{(3) State Update:}} 
        The governance mechanism updates:
        \begin{itemize}[leftmargin=30pt]
            \blockgap
            \item \textbf{Topic-level social state:}
            \[
                \Theta_k^t 
                = (1 - \lambda_s)\Theta_k^{t-1}
                + \lambda_s\Big[(r_k^t - r_k^{t-1})\,\overline{q}_k^t 
                - \gamma B_k^t\Big],
            \]
            where \(\lambda\) is the learning rate, \(\overline{q}_k^t\) is the supporter-quality term, and \(B_k^t\) is the anti-bubble penalty (defined below).    
            \blockgap
            \item \textbf{Agent-level credibility:}
            \[
                c_i^t 
                = c_i^{t-1}
                + \eta\big(\Theta_{a_i^t}^t - \Theta_{a_i^t}^{t-1}\big)\,
                  e^{-\kappa r_{a_i^t}^t},
            \]
            where \(\eta\) is a learning rate and \(\kappa\) implements an \emph{early-mover bonus}: agents who support a topic before it receives substantial resources (i.e. small \(r_{a_i^t}^t\)) experience a larger credibility update.

        \end{itemize}

    \addvspace{0.5em}
    \paragraph{\textbf{(4) Key Supporting Components:}}
    \begin{itemize}[leftmargin=30pt]
        \addvspace{0.5em}
        \item \textbf{Supporter Quality.}  
        CG measures the average credibility of agents supporting topic \(k\):
        \[
            \overline{q}_k^t
            = \frac{\sum_{i: a_i^t = k} c_i^{t-1}}
                   {\sum_{i: a_i^t = k} 1}.
        \]
        High-quality support contributes more to positive social signal updates.
    
        \addvspace{0.5em}
        \item \textbf{Anti-Bubble Penalty.}  
        CG penalizes rapid increases in support that lack credible backing:
        \[
            B_k^t
            = \sigma\!\big(\alpha(r_k^t - r_k^{t-1})\big)\,(1 - \overline{q}_k^t),
        \]
        suppressing low-quality bandwagons.
    
    
    \end{itemize}
    
    \addvspace{2.0em}
    \blocktitle{2.2.2 Web3-style Staking (WS)}
    WS models a “one-dollar, one-vote” token-based governance system. Agent influence is proportional to staked wealth, and the social signal \(\Theta_k^{t-1}\) reflects stake-weighted popularity. Unlike CG where influence weight is updated based on $\Delta$\(\Theta_k\), WS ties incentives directly to physical outcomes, adjusting an agent’s stake based on topic progress. This creates a direct feedback loop in which successful topics amplify the influence of their supporters.

    \paragraph{\textbf{(1) Perception:}}
        Agents observe the previous round’s stake-weighted popularity \(\Theta_k^{t-1} = r_k^{t-1}\) and noisy physical signal \(\tilde{\pi}_k^{\,t-1}\).
    
    \paragraph{\textbf{(2) Opinion Formation and Aggregation:}}
        Agents update beliefs and confidence \(\alpha_i^t\), selecting a topic \(a_i^t\).  
        Influence is stake-weighted:
        \[
            w_i^{t-1} = \alpha_i^t \text{bal}_i^{\,t-1}.
        \]
        The allocation is:
        \[
            r_k^t = 
            \frac{\sum_i w_i^{t-1}\mathbf{1}[a_i^t = k]}
                 {\sum_j w_j^{t-1}}.
        \]

    \addvspace{0.5em}
    \paragraph{\textbf{(3) State Update:}}
        \blockgap
        \begin{itemize}[leftmargin=30pt]
        \addvspace{0.5em}
            \item \textbf{Stake Update:}
            \[
                \text{bal}_i^t 
                = \text{bal}_i^{t-1} 
                + \gamma_s\, w_i^{t-1}\,\Delta \pi_{a_i^t}^t.
            \]
            where $\gamma$ is the staking reward rate, with the subscript $s$ indicating that it applies specifically to the Web3 Staking mechanism.
            \blockgap
            \item \textbf{Social Signal Update:}
            \[
                \Theta_k^t = r_k^t.
            \]
        \end{itemize}
        

    \addvspace{1.0em}
    \blocktitle{2.2.3 Social Media Upvote (SM)}
    SM represents a "one-person, one-vote" system. Influence weights are uniform regardless of past performance and agents vote purely based on their beliefs and observations. The social signal corresponds to raw popularity. No agent-level attributes are updated. 

        \paragraph{\textbf{(1) Perception:}}
        Agents observe last round’s raw popularity \(\Theta_k^{t-1} = r_k^{t-1}\) and the noisy physical signal \(\tilde{\pi}_k^{\,t-1}\).

        \addvspace{0.5em} 
        \paragraph{\textbf{(2) Opinion Formation and Aggregation:}} 
        Agents update beliefs and confidence, select a topic \(a_i^t\), and vote with equal influence:
        \[
            w_i^{t-1} = 1.
        \]
        The allocation is:
        \[
            r_k^t = \frac{1}{N}\sum_{i=1}^N \mathbf{1}[a_i^t = k].
        \]

        \addvspace{0.5em}
        \paragraph{\textbf{(3) State Update:}}
        \[
            \Theta_k^t = r_k^t.
        \]
        No credibility, stake, or other attributes are updated.


    \addvspace{1.0em}
    \blocktitle{2.2.4 No Governance (NG)}
    NG serves as a baseline with no social coordination, amplification, or feedback. Influence is uniform, no social signal is maintained, and no agent-level attributes are updated over time. This simulates an atomized society with no social component, where everyone simply observes signals from the physical world and deliberates on their own. 
    
        \paragraph{\textbf{(1) Perception:}}
        Agents do not receive a social signal; \(\Theta_k^{t-1} = 0\).  
        They observe only the noisy physical signal \(\tilde{\pi}_k^{\,t-1}\).
    
        \paragraph{\textbf{(2) Opinion Formation and Aggregation:}} 
        Agents vote with equal and fixed influence, the same as in SM. 
    
        \paragraph{\textbf{(3) State Update:}}
        No agent-level or topic-level feedback occurs; influence and (absent) social signals remain static.

    \subsection{Physical World Setups}
    In this work, we instantiate the Physical World as an academic setting with two scientific theories competing for collective resources. Their physical progress models how real-world scientific efforts unfold under different governance mechanisms.

    \addvspace{0.5em}
    \blocktitle{2.3.1 Topic Initialization}
    We consider two future-oriented scientific topics, \textbf{quantum physics (true Topic A)} and \textbf{neuromorphic physics (false Topic B)}, to avoid LLM omniscience, as current evidence supports both sides and it is inconclusive which is more valid. Truth is assigned arbitrarily to avoid pre-existing LLM biases. The initial physical progress for both topics is set to 0.

    \addvspace{0.5em}
    \blocktitle{2.3.2 Topic Progress Dynamics}
    \noindent
    Topic progress follows a nonlinear development trajectory. Let \(\pi_k^t\) denote the cumulative progress of topic \(k\) at round \(t\) and \(r_k^t\) the resource share allocated to topic \(k\) in that round. The increment \(\Delta \pi_k^{t}\) is given by:
    \[
    \Delta \pi_k^{t} =
    \begin{cases}
    \epsilon_k^{t}, 
    & \pi_k^{t-1} < \pi_{1}, \\[0.4em]
    \bigl(v_k + \gamma\, r_k^{t}\bigr)\!\left(1 - \dfrac{\pi_k^{t-1}}{M_k}\right) + \epsilon_k^{t}, 
    & \pi_{1} \le \pi_k^{t-1} < \pi_{2}, \\[0.4em]
    \gamma' r_k^{t}\!\left(1 - \dfrac{\pi_k^{t-1}}{M_k}\right) + \epsilon_k^{t}, 
    & \pi_k^{t-1} \ge \pi_{2}.
    \end{cases}
    \]
    This models an initial \textit{exploration} stage, an \textit{acceleration} stage, and a \textit{saturation} stage. Here, \(v_k\) is the intrinsic baseline velocity of topic \(k\), \(\gamma\) and \(\gamma'\) control how resources accelerate progress in different growth phases, \(M_k\) is the saturation limit, and \(\epsilon_k^t \sim \mathcal{N}(0,\sigma^2)\) captures environmental shocks.  

    \addvspace{0.5em}
    \blocktitle{2.3.3 Agent Population and Initial Conditions}
    We initialize \(N = 100\) agents with the following belief and epistemic profiles. The \(7{:}2{:}1\) split allows us to test whether the population can self-correct from a skewed initial state and to observe whether high-quality agents can rise in influence over time. 
    \begin{itemize}
        \item \textbf{Misaligned Majority (70 agents):} Start with a belief in the false Topic B and have moderate epistemic stability.
        \item \textbf{Truth-Aligned Minority (20 agents):} Start with a belief in the true Topic A and remain moderately influenceable.
        \item \textbf{High-Conviction Core (10 agents):} Start with a belief in the true Topic A and have very high epistemic stability, anchoring the system around ground truth.
    \end{itemize}
    \blockgap
    Unless otherwise specified, agent- and topic-level states are initialized as:
    \begin{itemize}
        \item \textbf{Credibility (CG):} \(c_i^{0} = 1\) for all agents.
        \item \textbf{Stake (WS):} \(\text{bal}_i^{0} = 1\) for all agents.
        \item \textbf{Social signal (CG):} \(\Theta_k^{0} = 0\) for both topics.
        \item \textbf{Confidence:} \(\alpha_i^{0} = 0.5\) for all agents.
    \end{itemize}

    \noindent
    These initial conditions ensure that all mechanisms start from a symmetric, low-information state: no topic has accumulated progress, no agent has privileged influence, the social signal has not yet formed, and all agents start with a moderately strong initial confidence level.

\subsection{Mapping CG to Real-World Examples}
\label{sec:mapping}

Table~\ref{tab:mapping} maps CG variables to concrete, time-indexed proxies in three data-rich examples (forecasting, peer review, and forums). CG only requires (i) a public social signal \(\Theta_k^t\) (and its change \(\Delta\Theta_k^t\)) and (ii) an outcome-linked signal \(\pi_k^t\) that arrives with noise and delay (often observed via a proxy \(\bar{\pi}_k^t\)).

\section{Experiment Results}
\label{sec:results}

\noindent
We present results in a hypothesis testing format. Each figure corresponds to a concrete claim about when Credibility Governance (CG) improves collective accuracy and why. Unless noted otherwise, each run lasts 30 rounds and we report mean trajectories over 10 trials with confidence bands.

\subsection{H1: CG accelerates truth convergence and improves shock recovery}
\label{sec:h1_system}

\begin{figure}
  \centering
  \includegraphics[width=\linewidth]{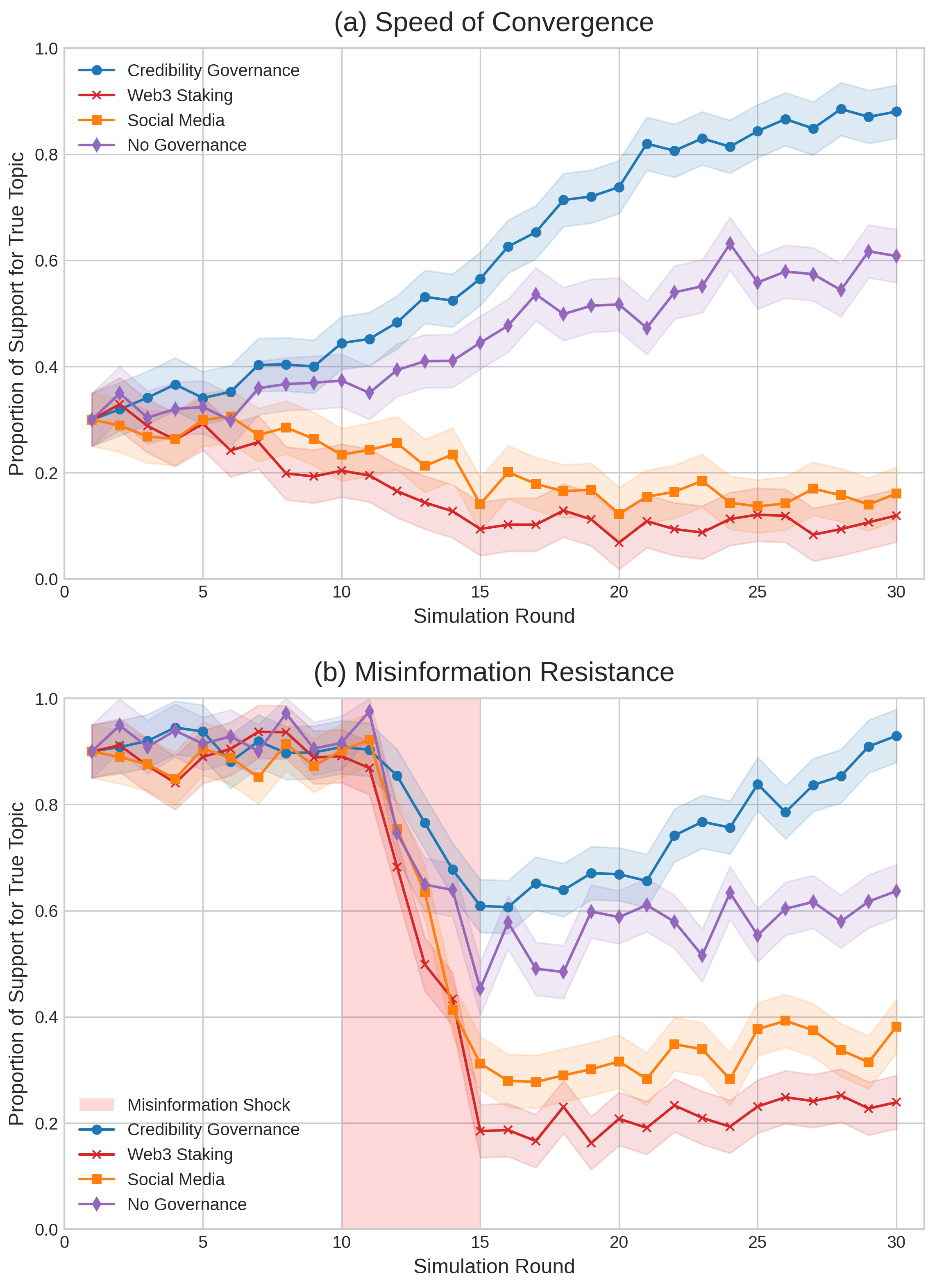}
  \caption{\textbf{H1, system dynamics.} Panel (a) shows convergence from an initially false-majority state. Panel (b) shows recovery after a misinformation shock (shaded window). CG reaches higher support for the true topic and recovers faster after shocks than WS, SM, and NG.}
  \label{fig:system_dynamics}
\end{figure}

\noindent
\textbf{Hypothesis (H1).} When truth is weakly observable, CG increases system-level accuracy by promoting correction over time, and recovers more effectively after misinformation shocks. \\
\textbf{Test.} We track the proportion of support for the true topic over rounds under (i) an initially false-majority state and (ii) an exogenous misinformation shock. \\
\textbf{Result.} In Fig.~\ref{fig:system_dynamics}a, CG steadily shifts collective belief toward the true topic and reaches a high-consensus regime, while WS and SM amplify early popularity and drift away from truth, and NG improves slowly. In Fig.~\ref{fig:system_dynamics}b, CG shows the strongest post-shock recovery and highest final accuracy, consistent with reduced lock-in and faster correction.


\subsection{H2: CG works by reallocating influence toward truth-aligned agents}
\label{sec:h2_mechanism}

\begin{figure}[h]
  \centering
  \includegraphics[width=\linewidth]{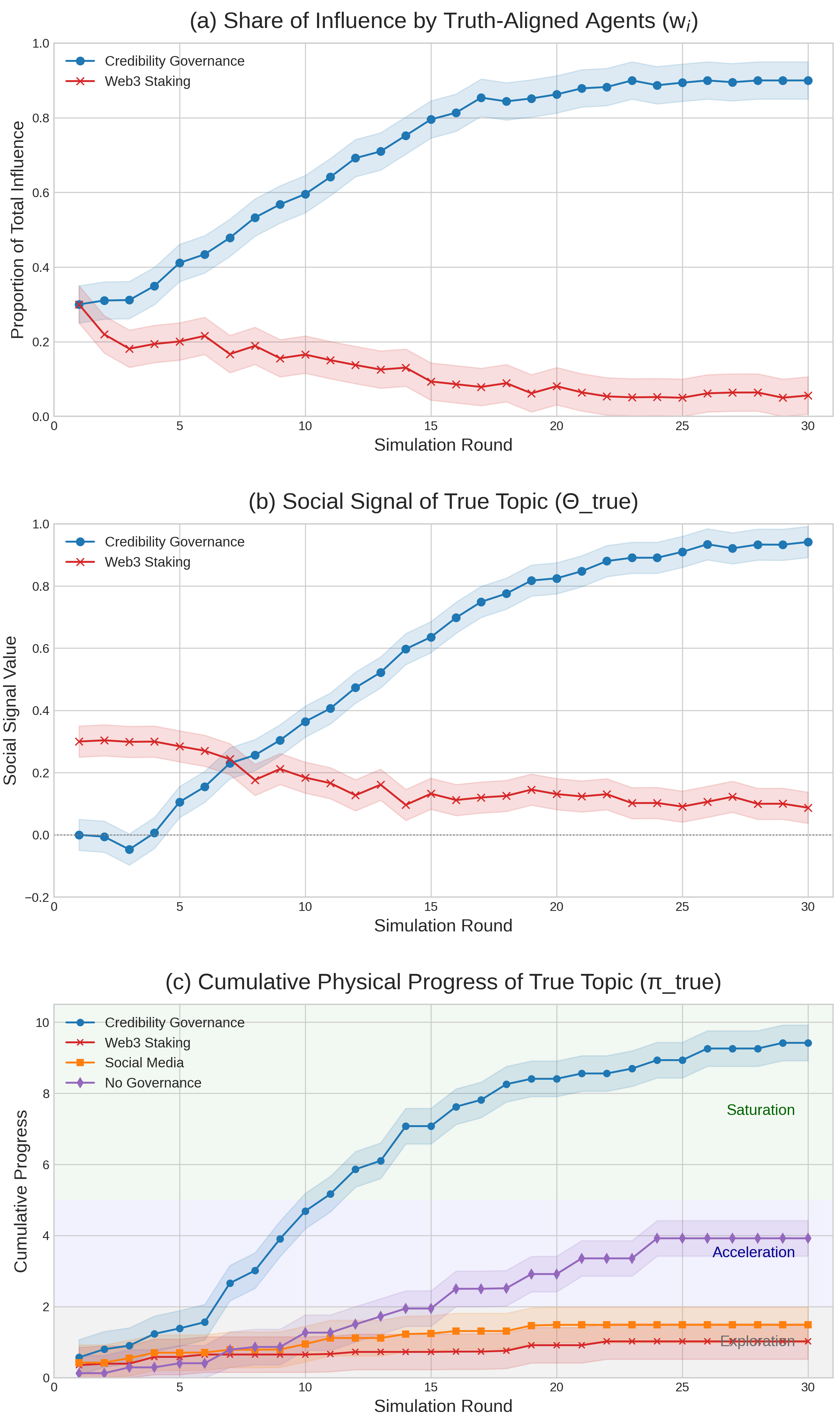}
  \caption{\textbf{H2, mechanism pathway.} Panel (a) tracks the share of influence held by truth-aligned agents. Panel (b) shows the social signal of the true topic $\Theta_{\text{true}}$. Panel (c) shows cumulative physical progress $\pi_{\text{true}}$, with background phases indicating exploration, acceleration, and saturation. CG reallocates influence toward truth-aligned agents, strengthens the social signal for the true topic, and compounds into faster physical progress.}
  \label{fig:mechanism_behavior}
\end{figure}

\noindent
\textbf{Hypothesis (H2).} CG improves system outcomes by shifting influence weights toward truth-aligned agents, which strengthens the social signal for the true topic and increases downstream physical progress. \\
\textbf{Test.} We measure (i) the share of total influence held by truth-aligned agents, (ii) the social signal of the true topic $\Theta_{\text{true}}$, and (iii) cumulative physical progress $\pi_{\text{true}}$. \\
\textbf{Result.} Fig.~\ref{fig:mechanism_behavior}a shows that CG concentrates influence among truth-aligned agents over time, unlike WS. This increases $\Theta_{\text{true}}$ (Fig.~\ref{fig:mechanism_behavior}b) and compounds into higher cumulative physical progress (Fig.~\ref{fig:mechanism_behavior}c), explaining the system-level gap observed in H1.


\subsection{H3: CG's core components are necessary, and $\Delta\Theta$ is the right reward basis}
\label{sec:h3_ablations}

\begin{figure}[h]
  \centering
  \includegraphics[width=0.90\linewidth]{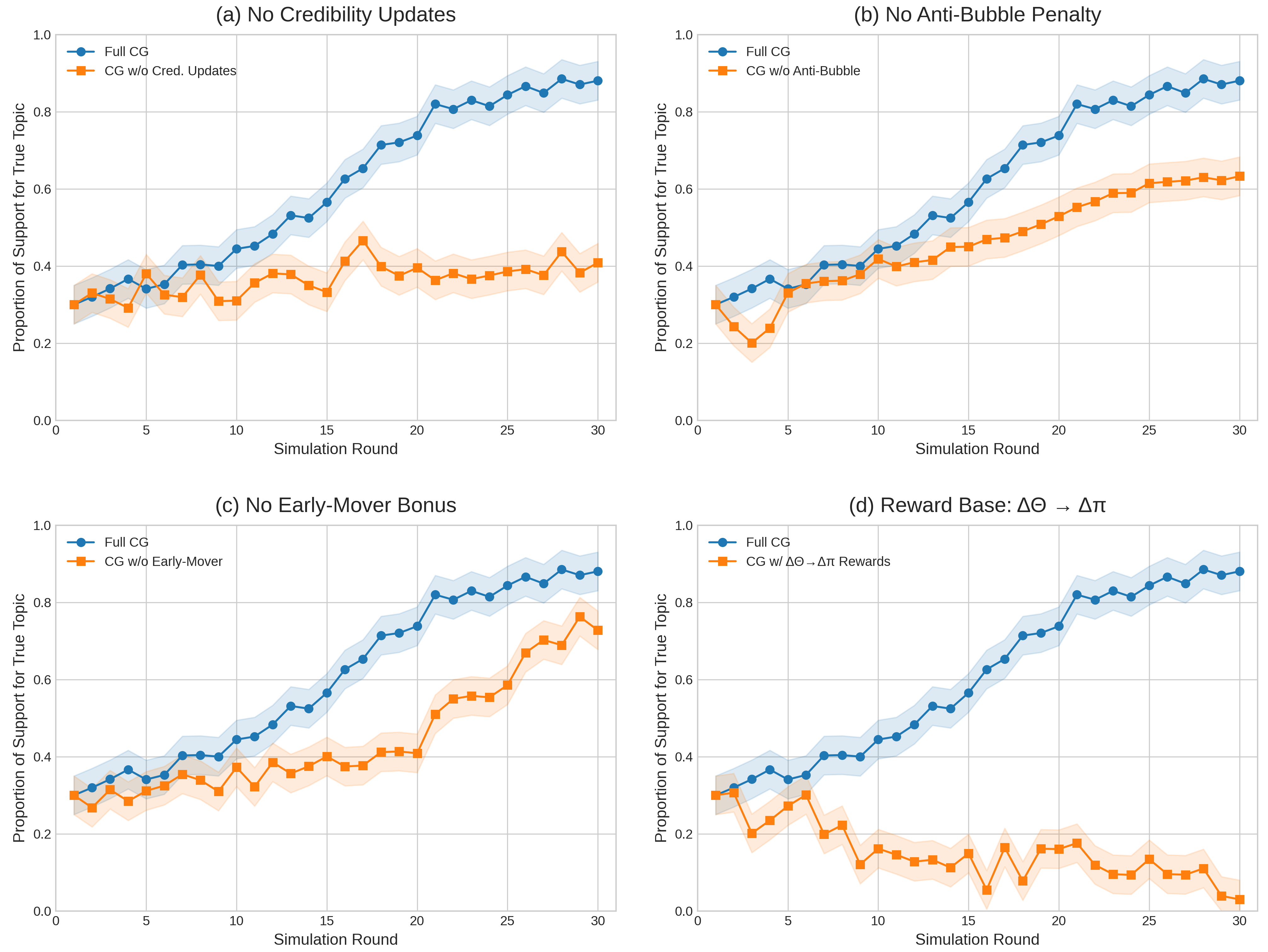}
  \caption{\textbf{H3, ablations.} Each panel compares support for the true topic in Full CG against an ablated variant: (a) no credibility updates, (b) no anti-bubble penalty, (c) no early-mover bonus, and (d) reward base $\Delta\Theta_k \rightarrow \Delta\pi_k$. Removing any component degrades performance, and replacing $\Delta\Theta$ with $\Delta\pi$ most strongly destabilizes convergence.}
  \label{fig:ablation_studies}
\end{figure}

\noindent
\textbf{Hypothesis (H3).} CG's performance depends on (i) credibility updates, (ii) an anti-bubble regularizer, (iii) an early-mover bonus that prevents premature lock-in, and critically (iv) using momentum in public evidence $\Delta\Theta_k$ rather than short-horizon physical proxies $\Delta\pi_k$ as the reward basis. \\
\textbf{Test.} We ablate each component and compare convergence trajectories against Full CG. \\
\textbf{Result.} Removing credibility updates sharply reduces convergence (Fig.~\ref{fig:ablation_studies}a). Removing the anti-bubble penalty increases cascade susceptibility and reduces final accuracy (Fig.~\ref{fig:ablation_studies}b). Removing the early-mover bonus slows the transition out of initial false-majority regimes (Fig.~\ref{fig:ablation_studies}c). Switching the reward basis to $\Delta\pi_k$ substantially weakens learning (Fig.~\ref{fig:ablation_studies}d), supporting $\Delta\Theta_k$ as a more reliable reward signal under noise and delay.

\subsection{Additional Hypotheses} 
We further test \textbf{H4} (CG is most beneficial at intermediate noise, by reducing lock-in and speeding correction), \textbf{H5} (Explicit credibility prompting provides a modest additional benefit), and \textbf{H6} (CG expands the safe operating region under adversarial manipulation). For brevity, these analyses are detailed in the Appendix (\ref{sec:h4_noise}--\ref{sec:h6_adversary}).

\section{Related Work}

\subsection{Truth Aggregation Mechanisms}

Systems for aggregating dispersed judgments underpin many web-based platforms, crowd-driven processes, and digital governance settings. Prior approaches correspond to the three baselines studied in this work: staking based mechanisms (as instantiated in Web3-style Staking, WS), vote based systems (as instantiated in Social Media Upvote, SM), and environments without explicit governance.

\addvspace{0.3em}
Staking-based mechanisms originate from decentralized coordination and blockchain governance frameworks \cite{buterin2014dpos, daian2020flashboys}. Participants lock resources behind a claim and receive rewards if their supported outcome proves correct \cite{hanson2006prediction}. Although these designs create incentives for accuracy, resource commitment does not reliably track epistemic quality. Because evidence production and amplification often depend on resources \cite{frye2021credence}, actors with greater resources can disproportionately shape outcomes independent of judgment reliability.

\addvspace{0.3em}
Vote-based and engagement-driven systems, widely used in social media and content ranking platforms \cite{leskovec2010predicting, muchnik2013social}, democratize participation but privilege visibility and momentum. Small early surges can cascade, allowing popularity to overshadow slower moving evidence. Reputation-based voting, common in peer review and trust-scoring environments \cite{resnick2000reputation, squires2021peerreview}, incorporates historical reliability, yet accumulated reputation can reduce responsiveness to new conditions and allow previously credible actors to retain influence after their judgments diverge from emerging evidence.

\addvspace{0.3em}
No-governance settings form a third class, in which agents observe noisy physical signals and deliberate independently. While simple and transparent, such systems remain sensitive to early noise and path dependence \cite{salganik2006experimental}, motivating mechanisms that better support collective self correction under uncertainty.

\subsection{Collective Misinformation Dynamics}

Work in collective epistemology shows that groups can misaggregate information when early signals are ambiguous or socially reinforced, producing herding and information cascades \cite{bikhchandani1992theory, banerjee1992simple}. These effects intensify under partial observability or uneven evidence quality \cite{golub2010naivete}, where competent individuals nonetheless inherit distorted beliefs.

\addvspace{0.3em}
Parallel research on misinformation dynamics demonstrates how visibility, repetition, and coordinated amplification enable weakly supported claims to diffuse faster than accurate but slower moving evidence \cite{vosoughi2018spread, pennycook2021psychology}. Platform ranking and feedback systems can further entrench these trajectories, locking communities into fragile consensus states \cite{bakshy2015exposure, delvicario2016echo}.

\addvspace{0.3em}
These findings underscore the need for governance frameworks that temper early amplification, identify reliable contributors, and maintain epistemic resilience when truth is only weakly observable. Our work speaks to this tradition by proposing a mechanism that operationalizes a central principle in social epistemology: collective accuracy depends on the community’s ability to evaluate the reliability of information sources \cite{goldman1999knowledge}. Credibility Governance implements this principle in a dynamic setting and tests it under conditions characterized by noisy, delayed, and uneven evidence.

\subsection{LLM-Based Social Simulation}
Agent-based models have long been used to study how individual level behaviors produce aggregate social outcomes \cite{epstein1996growing}. Traditional simulations rely on simplified rules or fixed parameters, limiting their ability to capture richer cognitive processes. Large language models have transformed this landscape by enabling agents capable of natural language interaction~\citep{he2025make, li2025beda, kang2020incorporating} and representation various value orientations~\citep{zhang2026bach, ziheng2026simple, kang-etal-2025-values} in open-ended social and even political scenarios~\citep{ziheng2025llm, smith2025evaluating, zhang2025policon, mao2025ibgp, park2023generative, argyle2023out}, weighing qualitative evidence, updating beliefs through deliberation, and articulating justifications. This shift has produced a growing body of literature using LLM agents to study collective behavior, coordination, and emergent social dynamics \cite{bratton2023model, chen2024agentverse, schramowski2023language}.

\addvspace{0.3em}
Our work builds on this paradigm by using LLM-based social simulations as an evaluation environment for Credibility Governance. In contrast to numerical opinion update models or rule-based simulations, LLM agents enable the study of governance mechanisms in settings where decisions are shaped by language-driven reasoning and context-sensitive interpretations of evidence. This allows us to analyze how influence and credibility evolve when truth is weakly observable and must be inferred from evolving, partially reliable signals.

\section{Discussion}
\label{sec:discussion}

Our experiments support a concrete scientific claim about collective learning under weakly observable truth signals.
When public evidence is noisy and partially contaminated, incentive and aggregation mechanisms that reward \emph{credible momentum} (changes in a filtered public-evidence proxy) can improve long-run accuracy by reallocating influence toward truth-aligned agents, reducing early lock-in, and accelerating error correction.

\subsection{Boundary conditions and where the mechanism can fail}
Our results also clarify boundary conditions.
First, the noise sweep shows a regime structure, CG provides the largest gains when signals are noisy yet still informative, while \textbf{too much noise breaks any mechanism} and all methods degrade toward chance-level performance.
Second, when the public-evidence proxy becomes systematically misaligned with ground truth, for example due to persistent contamination or delayed, biased observability, credibility updates may mis-allocate influence and slow correction.
Third, strong early coordination can transiently dominate momentum signals, increasing the risk of path-dependent outcomes even when the long-run mechanism is stabilizing.
These observations motivate reporting diagnostics alongside accuracy, because similar end accuracy can arise from different failure modes.

\subsection{Threat model limitations and mitigations}
Our adversarial evaluation covers representative strategic behaviors and two attacker implementations, but the threat model remains incomplete.
Two important threats are not yet tested.
\textbf{(i) Adaptive mimicry attacks:} attackers that dynamically adjust behavior to imitate credible trajectories, for example alternating between truthful and deceptive contributions to harvest credibility, then cashing it in during critical windows.
\textbf{(ii) Collusion and bribery attacks:} coordinated groups that exchange off-platform incentives, or use side payments to synchronize support shifts that resemble organic momentum.
Both threats target the core signal used by CG, credible momentum, and may reduce the separability between genuine evidence accumulation and manufactured dynamics.

Mitigations follow directly from these failure mechanisms.
One line is \textbf{credibility inertia and auditability}, add update friction, caps, or decay so that sudden credibility gains require sustained evidence, and incorporate lightweight audits on high-impact agents or bursts of momentum.
A second line is \textbf{cross-signal consistency checks}, require agreement across multiple observable channels, for example prediction accuracy, peer-assessed contributions, and longitudinal behavior, before large influence reallocations occur.
A third line is \textbf{robust aggregation under strategic manipulation}, integrate detection features such as burstiness, reciprocity, and network-based collusion signals into the public-evidence proxy.
These mitigations are compatible with the mechanism structure and can be evaluated within the same stress-test framework.

\subsection{Next validation on real traces}
A natural next step is validation on real traces, using the mapping section as an interface between the simulation variables and observed signals.
Concretely, we can instantiate $\Theta$ using domain-specific proxies, such as forecast accuracy in prediction settings, review helpfulness and calibration in peer review settings, or longitudinal post quality and community feedback in forums.
Then we can replay influence updates on historical traces, test counterfactual governance rules, and measure whether CG-style updates improve downstream outcomes such as agreement with ex post ground truth, reduced volatility, and faster recovery after misinformation bursts.
This trace-based validation would also enable stronger threat modeling, since adaptive and collusive behavior can be approximated from real temporal patterns and network structure, and the resulting observations can be used to refine the proxy definition of $\Theta$ and the update rules.

\section{Conclusion}
\label{sec:conclusion}

We studied how to design incentive and aggregation mechanisms that improve the accuracy and robustness of collective knowledge when truth is weakly observable and social signals are noisy or manipulated. We proposed Credibility Governance (CG), which reallocates influence using momentum in a public-evidence proxy, and evaluated it through system dynamics, ablations, diagnostic failure-mode metrics, noise stress tests, and adversarial threat models. Across regimes where evidence remains informative, CG reduces early lock-in and path dependence, shortens correction lag, and expands the stable operating region under adversarial pressure compared to popularity-driven and stake-amplified baselines, while clarifying a key boundary condition, when noise becomes extreme, all mechanisms degrade toward chance. Our mapping to real-world observable signals and the proposed trace-based validation plan provide a path to testing these claims on real information ecosystems and to tightening threat models with domain-grounded proxies.

\bibliographystyle{ACM-Reference-Format}
\bibliography{sample-base}


\begin{thebibliography}{39}


\ifx \showCODEN    \undefined \def \showCODEN     #1{\unskip}     \fi
\ifx \showISBNx    \undefined \def \showISBNx     #1{\unskip}     \fi
\ifx \showISBNxiii \undefined \def \showISBNxiii  #1{\unskip}     \fi
\ifx \showISSN     \undefined \def \showISSN      #1{\unskip}     \fi
\ifx \showLCCN     \undefined \def \showLCCN      #1{\unskip}     \fi
\ifx \shownote     \undefined \def \shownote      #1{#1}          \fi
\ifx \showarticletitle \undefined \def \showarticletitle #1{#1}   \fi
\ifx \showURL      \undefined \def \showURL       {\relax}        \fi
\providecommand\bibfield[2]{#2}
\providecommand\bibinfo[2]{#2}
\providecommand\natexlab[1]{#1}
\providecommand\showeprint[2][]{arXiv:#2}

\bibitem[Argyle et~al\mbox{.}(2023)]%
        {argyle2023out}
\bibfield{author}{\bibinfo{person}{Lisa~P. Argyle}, \bibinfo{person}{Ethan~C.
  Busby}, \bibinfo{person}{Nancy Fulda}, \bibinfo{person}{Joshua~R. Gubler},
  \bibinfo{person}{Christopher Rytting}, {and} \bibinfo{person}{David
  Wingate}.} \bibinfo{year}{2023}\natexlab{}.
\newblock \showarticletitle{Out of One, Many: Using Language Models to Simulate
  Human Samples}.
\newblock \bibinfo{journal}{\emph{Political Analysis}} \bibinfo{volume}{31},
  \bibinfo{number}{3} (\bibinfo{year}{2023}), \bibinfo{pages}{337--351}.
\newblock


\bibitem[Bakshy et~al\mbox{.}(2015a)]%
        {bakshy2015exposure}
\bibfield{author}{\bibinfo{person}{Eytan Bakshy}, \bibinfo{person}{Solomon
  Messing}, {and} \bibinfo{person}{Lada Adamic}.}
  \bibinfo{year}{2015}\natexlab{a}.
\newblock \showarticletitle{Exposure to ideologically diverse news and opinion
  on Facebook}.
\newblock \bibinfo{journal}{\emph{Science}} (\bibinfo{year}{2015}).
\newblock


\bibitem[Bakshy et~al\mbox{.}(2015b)]%
        {bakshy2015}
\bibfield{author}{\bibinfo{person}{Eytan Bakshy}, \bibinfo{person}{Solomon
  Messing}, {and} \bibinfo{person}{Lada~A Adamic}.}
  \bibinfo{year}{2015}\natexlab{b}.
\newblock \showarticletitle{Exposure to ideologically diverse news and opinion
  on Facebook}.
\newblock \bibinfo{journal}{\emph{Science}} \bibinfo{volume}{348},
  \bibinfo{number}{6239} (\bibinfo{year}{2015}), \bibinfo{pages}{1130--1132}.
\newblock


\bibitem[Banerjee(1992)]%
        {banerjee1992simple}
\bibfield{author}{\bibinfo{person}{Abhijit Banerjee}.}
  \bibinfo{year}{1992}\natexlab{}.
\newblock \showarticletitle{A simple model of herd behavior}.
\newblock \bibinfo{journal}{\emph{Quarterly Journal of Economics}}
  (\bibinfo{year}{1992}).
\newblock


\bibitem[Bikhchandani et~al\mbox{.}(1992)]%
        {bikhchandani1992theory}
\bibfield{author}{\bibinfo{person}{Sushil Bikhchandani}, \bibinfo{person}{David
  Hirshleifer}, {and} \bibinfo{person}{Ivo Welch}.}
  \bibinfo{year}{1992}\natexlab{}.
\newblock \showarticletitle{A theory of fads, fashion, custom, and cultural
  change as informational cascades}.
\newblock \bibinfo{journal}{\emph{Journal of Political Economy}}
  (\bibinfo{year}{1992}).
\newblock


\bibitem[Bratton et~al\mbox{.}(2023)]%
        {bratton2023model}
\bibfield{author}{\bibinfo{person}{Katie Bratton} {et~al\mbox{.}}}
  \bibinfo{year}{2023}\natexlab{}.
\newblock \showarticletitle{Modeling Emergent Social Behavior in Multi-Agent
  LLM Systems}. In \bibinfo{booktitle}{\emph{NeurIPS}}.
\newblock


\bibitem[Buterin(2014)]%
        {buterin2014dpos}
\bibfield{author}{\bibinfo{person}{Vitalik Buterin}.}
  \bibinfo{year}{2014}\natexlab{}.
\newblock \bibinfo{title}{On Stake and Consensus}.
\newblock \bibinfo{howpublished}{Ethereum Blog}.
\newblock


\bibitem[Chen et~al\mbox{.}(2024)]%
        {chen2024agentverse}
\bibfield{author}{\bibinfo{person}{Zhenyu Chen} {et~al\mbox{.}}}
  \bibinfo{year}{2024}\natexlab{}.
\newblock \showarticletitle{AgentVerse: A Flexible Framework for Multi-Agent
  LLM Simulation}. In \bibinfo{booktitle}{\emph{ICLR}}.
\newblock


\bibitem[Daian et~al\mbox{.}(2020)]%
        {daian2020flashboys}
\bibfield{author}{\bibinfo{person}{Philip Daian} {et~al\mbox{.}}}
  \bibinfo{year}{2020}\natexlab{}.
\newblock \showarticletitle{Flash Boys 2.0: Frontrunning, Transaction
  Reordering, and Consensus Instability}. In \bibinfo{booktitle}{\emph{IEEE
  S\&P}}.
\newblock


\bibitem[Del~Vicario et~al\mbox{.}(2016)]%
        {delvicario2016echo}
\bibfield{author}{\bibinfo{person}{Michela Del~Vicario} {et~al\mbox{.}}}
  \bibinfo{year}{2016}\natexlab{}.
\newblock \showarticletitle{Echo chambers in the age of misinformation}.
\newblock \bibinfo{journal}{\emph{PNAS}} (\bibinfo{year}{2016}).
\newblock


\bibitem[Epstein and Axtell(1996)]%
        {epstein1996growing}
\bibfield{author}{\bibinfo{person}{Joshua Epstein} {and}
  \bibinfo{person}{Robert Axtell}.} \bibinfo{year}{1996}\natexlab{}.
\newblock \bibinfo{booktitle}{\emph{Growing Artificial Societies}}.
\newblock \bibinfo{publisher}{MIT Press}.
\newblock


\bibitem[Frye(2021)]%
        {frye2021credence}
\bibfield{author}{\bibinfo{person}{Timothy Frye}.}
  \bibinfo{year}{2021}\natexlab{}.
\newblock \showarticletitle{Credence goods, signaling, and the link between
  money and trust}.
\newblock \bibinfo{journal}{\emph{Comparative Political Studies}}
  (\bibinfo{year}{2021}).
\newblock


\bibitem[Goldman(1999)]%
        {goldman1999knowledge}
\bibfield{author}{\bibinfo{person}{Alvin Goldman}.}
  \bibinfo{year}{1999}\natexlab{}.
\newblock \bibinfo{booktitle}{\emph{Knowledge in a Social World}}.
\newblock \bibinfo{publisher}{Oxford University Press}.
\newblock


\bibitem[Golub and Jackson(2010a)]%
        {golub2010}
\bibfield{author}{\bibinfo{person}{Benjamin Golub} {and}
  \bibinfo{person}{Matthew~O Jackson}.} \bibinfo{year}{2010}\natexlab{a}.
\newblock \showarticletitle{Naive learning in social networks and the wisdom of
  crowds}.
\newblock \bibinfo{journal}{\emph{American Economic Journal: Microeconomics}}
  \bibinfo{volume}{2}, \bibinfo{number}{1} (\bibinfo{year}{2010}),
  \bibinfo{pages}{112--149}.
\newblock


\bibitem[Golub and Jackson(2010b)]%
        {golub2010naivete}
\bibfield{author}{\bibinfo{person}{Benjamin Golub} {and}
  \bibinfo{person}{Matthew~O. Jackson}.} \bibinfo{year}{2010}\natexlab{b}.
\newblock \showarticletitle{Naive learning in social networks and the wisdom of
  crowds}.
\newblock \bibinfo{journal}{\emph{American Economic Journal: Microeconomics}}
  (\bibinfo{year}{2010}).
\newblock


\bibitem[Hanson(2006)]%
        {hanson2006prediction}
\bibfield{author}{\bibinfo{person}{Robin Hanson}.}
  \bibinfo{year}{2006}\natexlab{}.
\newblock \showarticletitle{Foul play in information markets}.
\newblock \bibinfo{journal}{\emph{Economic Journal}} (\bibinfo{year}{2006}).
\newblock


\bibitem[He et~al\mbox{.}(2025)]%
        {he2025make}
\bibfield{author}{\bibinfo{person}{Buwei He}, \bibinfo{person}{Yang Liu},
  \bibinfo{person}{Zhaowei Zhang}, \bibinfo{person}{Zixia Jia},
  \bibinfo{person}{Huijia Wu}, \bibinfo{person}{Zhaofeng He},
  \bibinfo{person}{Zilong Zheng}, {and} \bibinfo{person}{Yipeng Kang}.}
  \bibinfo{year}{2025}\natexlab{}.
\newblock \showarticletitle{Make an Offer They Can't Refuse: Grounding Bayesian
  Persuasion in Real-World Dialogues without Pre-Commitment}.
\newblock \bibinfo{journal}{\emph{arXiv preprint arXiv:2510.13387}}
  (\bibinfo{year}{2025}).
\newblock


\bibitem[Kang et~al\mbox{.}(2025)]%
        {kang-etal-2025-values}
\bibfield{author}{\bibinfo{person}{Yipeng Kang}, \bibinfo{person}{Junqi Wang},
  \bibinfo{person}{Yexin Li}, \bibinfo{person}{Mengmeng Wang},
  \bibinfo{person}{Wenming Tu}, \bibinfo{person}{Quansen Wang},
  \bibinfo{person}{Hengli Li}, \bibinfo{person}{Tingjun Wu},
  \bibinfo{person}{Xue Feng}, \bibinfo{person}{Fangwei Zhong}, {and}
  \bibinfo{person}{Zilong Zheng}.} \bibinfo{year}{2025}\natexlab{}.
\newblock \showarticletitle{Are the Values of {LLM}s Structurally Aligned with
  Humans? A Causal Perspective}. In \bibinfo{booktitle}{\emph{Findings of the
  Association for Computational Linguistics: ACL 2025}},
  \bibfield{editor}{\bibinfo{person}{Wanxiang Che}, \bibinfo{person}{Joyce
  Nabende}, \bibinfo{person}{Ekaterina Shutova}, {and}
  \bibinfo{person}{Mohammad~Taher Pilehvar}} (Eds.).
  \bibinfo{publisher}{Association for Computational Linguistics},
  \bibinfo{address}{Vienna, Austria}, \bibinfo{pages}{23147--23161}.
\newblock
\showISBNx{979-8-89176-256-5}
\href{https://doi.org/10.18653/v1/2025.findings-acl.1188}{doi:\nolinkurl{10.18653/v1/2025.findings-acl.1188}}


\bibitem[Kang et~al\mbox{.}(2020)]%
        {kang2020incorporating}
\bibfield{author}{\bibinfo{person}{Yipeng Kang}, \bibinfo{person}{Tonghan
  Wang}, {and} \bibinfo{person}{Gerard de Melo}.}
  \bibinfo{year}{2020}\natexlab{}.
\newblock \showarticletitle{Incorporating pragmatic reasoning communication
  into emergent language}.
\newblock \bibinfo{journal}{\emph{Advances in neural information processing
  systems}}  \bibinfo{volume}{33} (\bibinfo{year}{2020}),
  \bibinfo{pages}{10348--10359}.
\newblock


\bibitem[Kleinberg and Easley(2022)]%
        {kleinberg2022trust}
\bibfield{author}{\bibinfo{person}{Jon Kleinberg} {and} \bibinfo{person}{David
  Easley}.} \bibinfo{year}{2022}\natexlab{}.
\newblock \showarticletitle{Trust in a Complex World: A Network Theory of
  Social Capital}.
\newblock \bibinfo{journal}{\emph{Commun. ACM}} (\bibinfo{year}{2022}).
\newblock


\bibitem[Leskovec et~al\mbox{.}(2010)]%
        {leskovec2010predicting}
\bibfield{author}{\bibinfo{person}{Jure Leskovec}, \bibinfo{person}{Daniel
  Huttenlocher}, {and} \bibinfo{person}{Jon Kleinberg}.}
  \bibinfo{year}{2010}\natexlab{}.
\newblock \showarticletitle{Predicting Positive and Negative Links in Online
  Social Networks}. In \bibinfo{booktitle}{\emph{WWW}}.
\newblock


\bibitem[Li et~al\mbox{.}(2025)]%
        {li2025beda}
\bibfield{author}{\bibinfo{person}{Hengli Li}, \bibinfo{person}{Zhaoxin Yu},
  \bibinfo{person}{Qi Shen}, \bibinfo{person}{Chenxi Li},
  \bibinfo{person}{Mengmeng Wang}, \bibinfo{person}{Tinglang Wu},
  \bibinfo{person}{Yipeng Kang}, \bibinfo{person}{Yuxuan Wang},
  \bibinfo{person}{Song-Chun Zhu}, \bibinfo{person}{Zixia Jia},
  {et~al\mbox{.}}} \bibinfo{year}{2025}\natexlab{}.
\newblock \showarticletitle{BEDA: Belief Estimation as Probabilistic
  Constraints for Performing Strategic Dialogue Acts}.
\newblock \bibinfo{journal}{\emph{arXiv preprint arXiv:2512.24885}}
  (\bibinfo{year}{2025}).
\newblock


\bibitem[Lorenz et~al\mbox{.}(2011)]%
        {lorenz2011}
\bibfield{author}{\bibinfo{person}{Jan Lorenz}, \bibinfo{person}{Heiko Rauhut},
  \bibinfo{person}{Frank Schweitzer}, {and} \bibinfo{person}{Dirk Helbing}.}
  \bibinfo{year}{2011}\natexlab{}.
\newblock \showarticletitle{How social influence can undermine the wisdom of
  crowd effect}.
\newblock \bibinfo{journal}{\emph{PNAS}} \bibinfo{volume}{108},
  \bibinfo{number}{22} (\bibinfo{year}{2011}), \bibinfo{pages}{9020--9025}.
\newblock


\bibitem[Mao et~al\mbox{.}(2025)]%
        {mao2025ibgp}
\bibfield{author}{\bibinfo{person}{Yihuan Mao}, \bibinfo{person}{Yipeng Kang},
  \bibinfo{person}{Peilun Li}, \bibinfo{person}{Wei Xu}, {and}
  \bibinfo{person}{Chongjie Zhang}.} \bibinfo{year}{2025}\natexlab{}.
\newblock \showarticletitle{Ibgp: Imperfect byzantine generals problem for
  zero-shot robustness in communicative multi-agent systems}. In
  \bibinfo{booktitle}{\emph{International Conference on Artificial General
  Intelligence}}. Springer, \bibinfo{pages}{421--432}.
\newblock


\bibitem[Muchnik et~al\mbox{.}(2013a)]%
        {muchnik2013social}
\bibfield{author}{\bibinfo{person}{Lev Muchnik}, \bibinfo{person}{Sinan Aral},
  {and} \bibinfo{person}{Sean Taylor}.} \bibinfo{year}{2013}\natexlab{a}.
\newblock \showarticletitle{Social influence bias: A randomized experiment}.
\newblock \bibinfo{journal}{\emph{Science}} (\bibinfo{year}{2013}).
\newblock


\bibitem[Muchnik et~al\mbox{.}(2013b)]%
        {muchnik2013}
\bibfield{author}{\bibinfo{person}{Lev Muchnik}, \bibinfo{person}{Sinan Aral},
  {and} \bibinfo{person}{Sean~J Taylor}.} \bibinfo{year}{2013}\natexlab{b}.
\newblock \showarticletitle{Social influence bias: A randomized experiment}.
\newblock \bibinfo{journal}{\emph{Science}} \bibinfo{volume}{341},
  \bibinfo{number}{6146} (\bibinfo{year}{2013}), \bibinfo{pages}{647--651}.
\newblock


\bibitem[Park et~al\mbox{.}(2023)]%
        {park2023generative}
\bibfield{author}{\bibinfo{person}{Joon~Sung Park}, \bibinfo{person}{Joseph
  O'Brien}, \bibinfo{person}{Carrie~Jun Cai}, \bibinfo{person}{Meredith~Ringel
  Morris}, \bibinfo{person}{Percy Liang}, {and} \bibinfo{person}{Michael~S
  Bernstein}.} \bibinfo{year}{2023}\natexlab{}.
\newblock \showarticletitle{Generative agents: Interactive simulacra of human
  behavior}. In \bibinfo{booktitle}{\emph{Proceedings of the 36th annual acm
  symposium on user interface software and technology}}.
  \bibinfo{pages}{1--22}.
\newblock


\bibitem[Pennycook and Rand(2021)]%
        {pennycook2021psychology}
\bibfield{author}{\bibinfo{person}{Gordon Pennycook} {and}
  \bibinfo{person}{David Rand}.} \bibinfo{year}{2021}\natexlab{}.
\newblock \showarticletitle{The psychology of fake news}.
\newblock \bibinfo{journal}{\emph{Trends in Cognitive Sciences}}
  (\bibinfo{year}{2021}).
\newblock


\bibitem[Resnick and Zeckhauser(2000)]%
        {resnick2000reputation}
\bibfield{author}{\bibinfo{person}{Paul Resnick} {and} \bibinfo{person}{Richard
  Zeckhauser}.} \bibinfo{year}{2000}\natexlab{}.
\newblock \showarticletitle{Reputation systems}.
\newblock \bibinfo{journal}{\emph{Commun. ACM}} (\bibinfo{year}{2000}).
\newblock


\bibitem[Salganik et~al\mbox{.}(2006)]%
        {salganik2006experimental}
\bibfield{author}{\bibinfo{person}{Matthew Salganik}, \bibinfo{person}{Peter
  Dodds}, {and} \bibinfo{person}{Duncan Watts}.}
  \bibinfo{year}{2006}\natexlab{}.
\newblock \showarticletitle{Experimental study of inequality and
  unpredictability in an artificial cultural market}.
\newblock \bibinfo{journal}{\emph{Science}} (\bibinfo{year}{2006}).
\newblock


\bibitem[Schramowski et~al\mbox{.}(2023)]%
        {schramowski2023language}
\bibfield{author}{\bibinfo{person}{Patrick Schramowski} {et~al\mbox{.}}}
  \bibinfo{year}{2023}\natexlab{}.
\newblock \showarticletitle{Language Models as Simulated Societies}.
\newblock \bibinfo{journal}{\emph{Scientific Reports}} (\bibinfo{year}{2023}).
\newblock


\bibitem[Smith et~al\mbox{.}(2025)]%
        {smith2025evaluating}
\bibfield{author}{\bibinfo{person}{Chandler Smith}, \bibinfo{person}{Marwa
  Abdulhai}, \bibinfo{person}{Manfred Diaz}, \bibinfo{person}{Marko Tesic},
  \bibinfo{person}{Rakshit~S Trivedi}, \bibinfo{person}{Alexander~Sasha
  Vezhnevets}, \bibinfo{person}{Lewis Hammond}, \bibinfo{person}{Jesse
  Clifton}, \bibinfo{person}{Minsuk Chang}, \bibinfo{person}{Edgar~A
  Du{\'e}{\~n}ez-Guzm{\'a}n}, {et~al\mbox{.}}} \bibinfo{year}{2025}\natexlab{}.
\newblock \showarticletitle{Evaluating generalization capabilities of LLM-based
  agents in mixed-motive scenarios using concordia}.
\newblock \bibinfo{journal}{\emph{arXiv preprint arXiv:2512.03318}}
  (\bibinfo{year}{2025}).
\newblock


\bibitem[Squires et~al\mbox{.}(2021)]%
        {squires2021peerreview}
\bibfield{author}{\bibinfo{person}{Ethan Squires} {et~al\mbox{.}}}
  \bibinfo{year}{2021}\natexlab{}.
\newblock \showarticletitle{How Do Peer-Review Rating Systems Influence
  Scientific Consensus?}
\newblock \bibinfo{journal}{\emph{Science Advances}} (\bibinfo{year}{2021}).
\newblock


\bibitem[Sunstein(2006)]%
        {sunstein2006}
\bibfield{author}{\bibinfo{person}{Cass~R Sunstein}.}
  \bibinfo{year}{2006}\natexlab{}.
\newblock \bibinfo{booktitle}{\emph{Infotopia: How Many Minds Produce
  Knowledge}}.
\newblock \bibinfo{publisher}{Oxford University Press}.
\newblock


\bibitem[Vosoughi et~al\mbox{.}(2018)]%
        {vosoughi2018spread}
\bibfield{author}{\bibinfo{person}{Soroush Vosoughi}, \bibinfo{person}{Deb
  Roy}, {and} \bibinfo{person}{Sinan Aral}.} \bibinfo{year}{2018}\natexlab{}.
\newblock \showarticletitle{The spread of true and false news online}.
\newblock \bibinfo{journal}{\emph{Science}} (\bibinfo{year}{2018}).
\newblock


\bibitem[Zhang et~al\mbox{.}(2026)]%
        {zhang2026bach}
\bibfield{author}{\bibinfo{person}{Junyu Zhang}, \bibinfo{person}{Yipeng Kang},
  \bibinfo{person}{Jiong Guo}, \bibinfo{person}{Jiayu Zhan}, {and}
  \bibinfo{person}{Junqi Wang}.} \bibinfo{year}{2026}\natexlab{}.
\newblock \showarticletitle{BACH-V: Bridging Abstract and Concrete Human-Values
  in Large Language Models}.
\newblock \bibinfo{journal}{\emph{arXiv preprint arXiv:2601.14007}}
  (\bibinfo{year}{2026}).
\newblock


\bibitem[Zhang et~al\mbox{.}(2025)]%
        {zhang2025policon}
\bibfield{author}{\bibinfo{person}{Zhaowei Zhang}, \bibinfo{person}{Xiaobo
  Wang}, \bibinfo{person}{Minghua Yi}, \bibinfo{person}{Mengmeng Wang},
  \bibinfo{person}{Fengshuo Bai}, \bibinfo{person}{Zilong Zheng},
  \bibinfo{person}{Yipeng Kang}, {and} \bibinfo{person}{Yaodong Yang}.}
  \bibinfo{year}{2025}\natexlab{}.
\newblock \showarticletitle{PoliCon: Evaluating LLMs on Achieving Diverse
  Political Consensus Objectives}.
\newblock \bibinfo{journal}{\emph{arXiv preprint arXiv:2505.19558}}
  (\bibinfo{year}{2025}).
\newblock


\bibitem[Ziheng et~al\mbox{.}(2026)]%
        {ziheng2026simple}
\bibfield{author}{\bibinfo{person}{Zhou Ziheng}, \bibinfo{person}{Jiakun Ding},
  \bibinfo{person}{Zhaowei Zhang}, \bibinfo{person}{Ruosen Gao},
  \bibinfo{person}{Yingnian Wu}, \bibinfo{person}{Demetri Terzopoulos},
  \bibinfo{person}{Yipeng Kang}, \bibinfo{person}{Fangwei Zhong}, {and}
  \bibinfo{person}{Junqi Wang}.} \bibinfo{year}{2026}\natexlab{}.
\newblock \showarticletitle{Simple Role Assignment is Extraordinarily Effective
  for Safety Alignment}.
\newblock \bibinfo{journal}{\emph{arXiv preprint arXiv:2602.00061}}
  (\bibinfo{year}{2026}).
\newblock


\bibitem[Ziheng et~al\mbox{.}(2025)]%
        {ziheng2025llm}
\bibfield{author}{\bibinfo{person}{Zhou Ziheng}, \bibinfo{person}{Huacong
  Tang}, \bibinfo{person}{Mingjie Bi}, \bibinfo{person}{Yipeng Kang},
  \bibinfo{person}{Wanying He}, \bibinfo{person}{Fang Sun},
  \bibinfo{person}{Yizhou Sun}, \bibinfo{person}{Ying~Nian Wu},
  \bibinfo{person}{Demetri Terzopoulos}, {and} \bibinfo{person}{Fangwei
  Zhong}.} \bibinfo{year}{2025}\natexlab{}.
\newblock \showarticletitle{An LLM-based Agent Simulation Approach to Study
  Moral Evolution}.
\newblock \bibinfo{journal}{\emph{arXiv preprint arXiv:2509.17703}}
  (\bibinfo{year}{2025}).
\newblock


\end{thebibliography}
\appendix
\section{Key Parameters}
    
        \begin{table}
            \centering
            \label{tab:exp_params}
            \renewcommand{\arraystretch}{1.2}
            \begin{tabular}{l@{\hskip 0.8pt}c c p{2.0cm}}
            \toprule
            \textbf{Parameter} & \textbf{Symbol} & \textbf{Value} & \textbf{Scope / Description} \\
            \midrule
        
            \multicolumn{4}{l}{\textit{\textbf{Physical Progress Dynamics}}} \\
            \multicolumn{4}{l}{%
              \(\displaystyle
              \Delta \pi_k^{t} =
              \begin{cases}
              \epsilon_k^{t}, 
              & \pi_k^{t-1} < \pi_{1}, \\[0.4em]
              \bigl(v_k + \gamma\, r_k^{t}\bigr)\!\left(1 - \dfrac{\pi_k^{t-1}}{M_k}\right) + \epsilon_k^{t}, 
              & \pi_{1} \le \pi_k^{t-1} < \pi_{2}, \\[0.4em]
              \gamma' r_k^{t}\!\left(1 - \dfrac{\pi_k^{t-1}}{M_k}\right) + \epsilon_k^{t}, 
              & \pi_k^{t-1} \ge \pi_{2}.
              \end{cases}
              \)} \\
            Environmental Noise & \(\sigma\) & 0.1 & std. dev. of \(\epsilon_k^t\) \\
            Baseline Velocities & \(v_{\text{true}}, v_{\text{false}}\) & 1.0, 0.8 & intrinsic growth \\
            Acceleration Factor & \(\gamma\) & 0.8 & mid-stage resource impact \\
            Saturation Limits & \(M_{\text{true}}, M_{\text{false}}\) & 10, 8 & max progress \\
            Late-Stage Factor & \(\gamma'\) & 0.4 & late-stage resource impact \\
            Stage Thresholds & \(\pi_1, \pi_2\) & 2.0, 5.0 & phase transition points \\
            \midrule
        
            \multicolumn{4}{l}{\textit{\textbf{Agent Perception Model}}} \\
            \multicolumn{4}{l}{%
              \(\displaystyle
              \tilde{\pi}_k^t = \pi_k^{t-1} + \rho\,\Theta_k^{t-1} + \xi_k^t
              \)} \\
            Social Contamination & \(\rho\) & 0.2 & strength of social shaping \\
            Observation Noise & \(\xi_k^t\) & 0.05 & std. dev. of perception noise \\
            \midrule
        
            \multicolumn{4}{l}{\textit{\textbf{Credibility Governance (CG) Specific}}} \\
        
            \multicolumn{4}{l}{%
              \(\displaystyle
              w_i^{t-1} = \alpha_i^t e^{\lambda c_i^{t-1}}
              \)} \\
            Influence Concentration & \(\lambda\) & 2.0 & exponential influence scaling \\
        
            \multicolumn{4}{l}{%
              \(\displaystyle
              \Theta_k^t 
              = (1 - \lambda_s)\Theta_k^{t-1}
                + \lambda_s\Big[(r_k^t - r_k^{t-1})\,\overline{q}_k^t 
                - \gamma B_k^t\Big]
              \)} \\
            Social-Signal Smoothing & \(\lambda_s\) & 0.4 & smoothing of social signal update \\
        
            \multicolumn{4}{l}{%
              \(\displaystyle
              c_i^t 
              = c_i^{t-1}
                + \eta\big(\Theta_{a_i^t}^t - \Theta_{a_i^t}^{t-1}\big)\,
                  e^{-\kappa r_{a_i^t}^t}
              \)} \\
            Credibility Learning Rate  & \(\eta\) & 0.1 & rate of credibility change \\
            Early-Mover Advantage & \(\kappa\) & 0.5 & early-mover bonus factor \\
        
            \multicolumn{4}{l}{%
              \(\displaystyle
              \overline{q}_k^t 
              = \frac{\sum_{i: a_i^t = k} c_i^{t-1}}
                     {\sum_{i: a_i^t = k} 1},
              \quad
              B_k^t 
              = \sigma\!\big(\alpha(r_k^t - r_k^{t-1})\big)\,\big(1 - \overline{q}_k^t\big)
              \)} \\
            Inconsistency Penalty & \(\beta\) & 0.3 & penalty for switching \\
            \midrule
        
            \multicolumn{4}{l}{\textit{\textbf{Web3 Staking (WS) Specific}}} \\
            \multicolumn{4}{l}{%
              \(\displaystyle
              w_i^{t-1} = \alpha_i^t\,\text{bal}_i^{t-1},
              \quad
              \text{bal}_i^t 
              = \text{bal}_i^{t-1} + \gamma_s w_i^{t-1}\,\Delta\pi_{a_i^t}^t
              \)} \\
            Staking Reward Rate & \(\gamma_s\) & 0.1 & proportion of progress paid as stake reward \\
            \bottomrule
            \end{tabular}
        \end{table}
  
        \noindent
        The specific parameter values used in our experiments are summarized in Table A. We group them by scope (global, perception-related, and mechanism-specific).
    
        \paragraph{Diagnostics and failure-mode metrics.}
        To make robustness results interpretable, we report a small set of diagnostics that capture classical failure modes in collective learning under noisy social signals.
    
        \textbf{Early lock-in.}
        We measure the probability that the system commits to the wrong topic early and does not recover.
        Concretely, an episode is counted as an early lock-in failure if the false topic reaches a high-support threshold $\tau$ within the first $T_{\mathrm{early}}$ rounds and the final support of the true topic remains below $\tau$ at the end of the run.
        We report early lock-in as the fraction of episodes that satisfy this criterion.
    
        \textbf{Path dependence.}
        We quantify how strongly the final outcome depends on early random fluctuations.
        Concretely, we run paired trials that share the same initial conditions but differ only in stochastic perception noise, and compute the dispersion of final support for the true topic, e.g.,
        $\mathrm{PD}=\mathrm{Var}\!\left[s_{\mathrm{true}}^{(r)}(T)\right]$ across replicate runs $r$.
        Higher dispersion indicates stronger path dependence.
    
        \textbf{Correction lag.}
        We measure how quickly the system corrects after evidence flips against the currently dominant topic.
        Let $t^\star$ denote the first round where the true topic becomes favorable in public evidence (or equivalently where the evidence proxy $\Theta_{\mathrm{true}}$ surpasses $\Theta_{\mathrm{false}}$ by margin $\delta$).
        Correction lag is defined as the number of rounds until the support for the true topic crosses a target threshold $\tau$, i.e.,
        $\mathrm{Lag}=\min\{t \ge t^\star : s_{\mathrm{true}}(t)\ge \tau\}-t^\star$,
        and we normalize it to $[0,1]$ for plotting.
    
        \textbf{Influence concentration.}
        We measure whether influence mass collapses onto a small set of agents.
        Let $w_i(t)$ be agent $i$'s normalized influence weight at round $t$ (summing to 1).
        We report the (time-averaged) Herfindahl index,
        $\mathrm{IC}=\frac{1}{T}\sum_{t=1}^{T}\sum_{i=1}^{N} w_i(t)^2$,
        where larger values indicate greater concentration.
    
        \paragraph{Baselines and fairness upgrades.}
        We compare CG against three governance baselines, Web3 staking (WS), social media upvotes (SM), and no governance (NG), under the same simulation budget and evaluation protocol.
        To ensure a fair stress test, our WS implementation includes the same practical frictions that arise in noisy information ecosystems, namely \textbf{(i) noisy stake updates} and \textbf{(ii) delayed stake visibility}, so that stake-weighting is not evaluated under unrealistically clean signals.
        We also include an explicit \textbf{credibility prompting} control, which isolates potential \emph{instructional bias} by changing only the agent instructions, while keeping the mechanism and environment fixed.
        Across all baselines and CG variants, we tuned hyperparameters with the same search budget and selected configurations using the same validation criterion (i.e., identical tuning budget and selection metric).

    \section{Parameter Sensitivity Analysis}
    \subsection{Sensitivity to Population Size (N)}
    \begin{figure}
        \centering
        \includegraphics[width=1.0\linewidth]{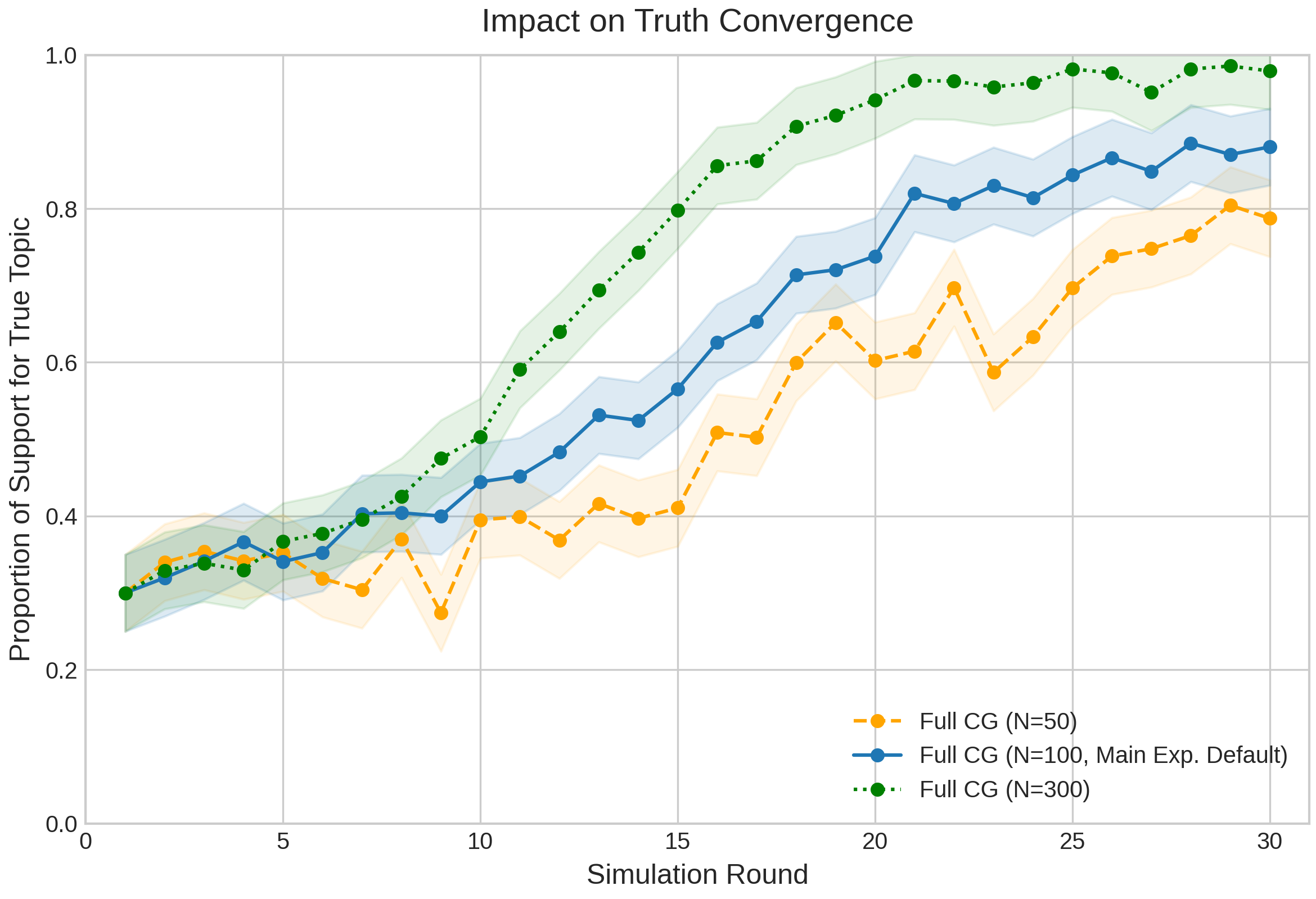}
        \caption{Sensitivity to Population Size. Comparison of CG's convergence for N=50, N=100 (Main Experiment), and N=300.}
        \label{fig:population_size}
    \end{figure}

    \noindent
    Figure~\ref{fig:population_size} shows how population size (N) affects CG’s convergence. Larger populations (N=300) converge faster and more smoothly than N=100 and especially N=50. With more agents, the social signal produced by the LLM-based population becomes more stable: individual stochasticity in model generations is averaged out, and credibility updates reflect consistent patterns rather than noise. This stronger aggregate signal enables CG to identify reliable agents more quickly and reinforce the correct topic. Smaller populations, by contrast, are more exposed to randomness from individual LLM outputs, making the social signal less stable and slowing convergence.

    \subsection{Sensitivity to Initial Opinion Distribution}

    \begin{figure}
        \centering
        \includegraphics[width=1.0\linewidth]{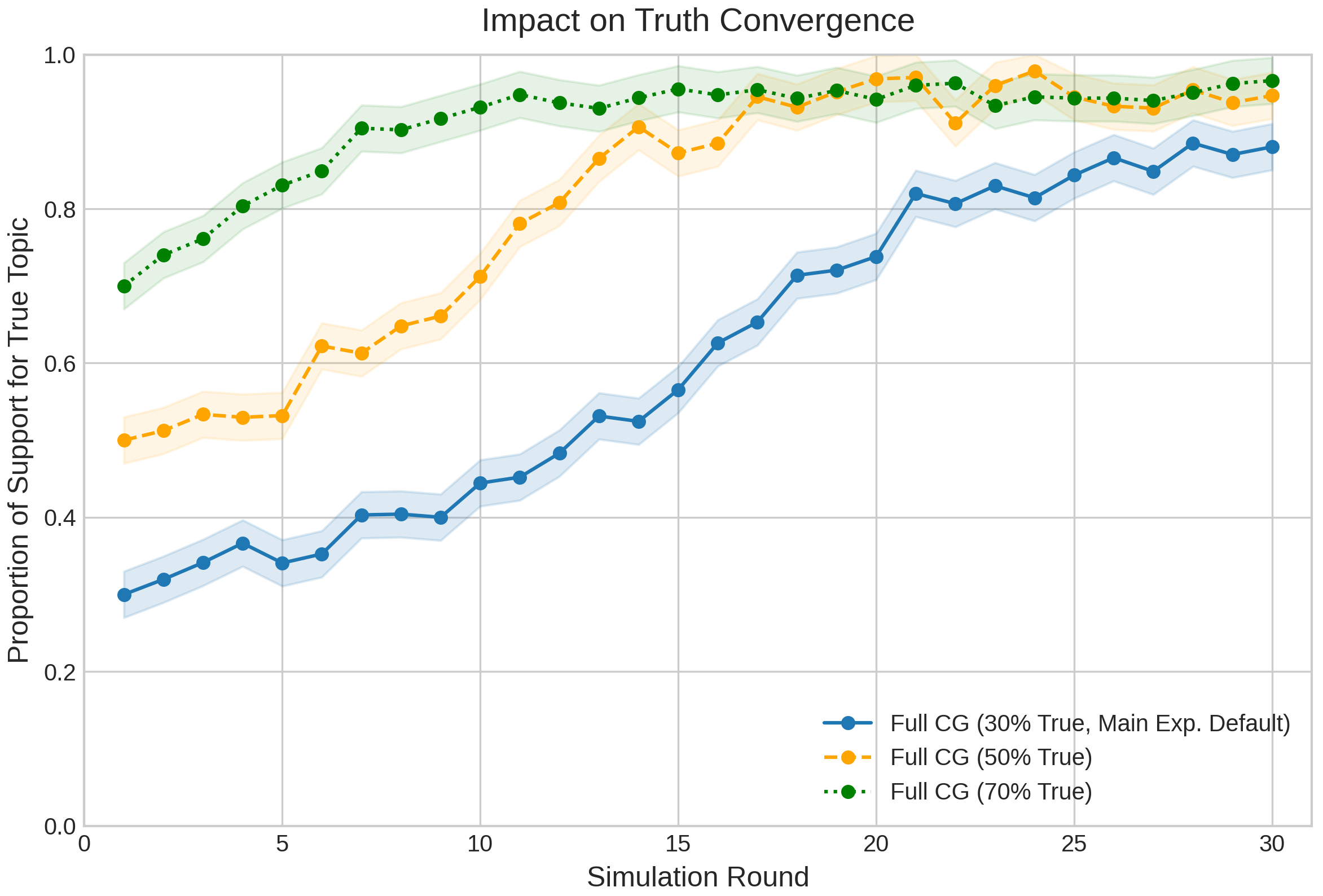}
        \caption{Sensitivity to Initial Opinion Distribution. Comparison of CG's convergence for 30\%, 50\%, and 70\% initial truth-aligned agents.}
        \label{fig:appendix:initial_opinion}
    \end{figure}

    \noindent
    Figure~\ref{fig:appendix:initial_opinion} shows how the initial opinion distribution affects CG’s convergence. As the share of truth-aligned agents increases (30\%, 50\%, 70\%), convergence becomes faster and more stable. When truth begins as the majority, CG reinforces it almost immediately. This is because the true topic produces more consistent improvement signals in the early rounds, giving truth-aligned agents slightly positive credibility updates from the start. As these small advantages accumulate, the average credibility of truth supporters rises within the first few rounds. Once their mean credibility exceeds that of false supporters, the anti-bubble term for the true topic becomes negligible, since it scales with $(1 - \bar{q}_{\text{true}}^t)$. Thus, a truth-majority receives a steady increase rather than a penalty. Overall, CG not only overcomes unfavorable initial conditions but also rapidly consolidates already truth-aligned populations.

    \subsection{Sensitivity to Baseline Velocity Gap} 
    \label{sec:appendix:velocity_gap}
    \begin{figure}
        \centering
        \includegraphics[width=1\linewidth]{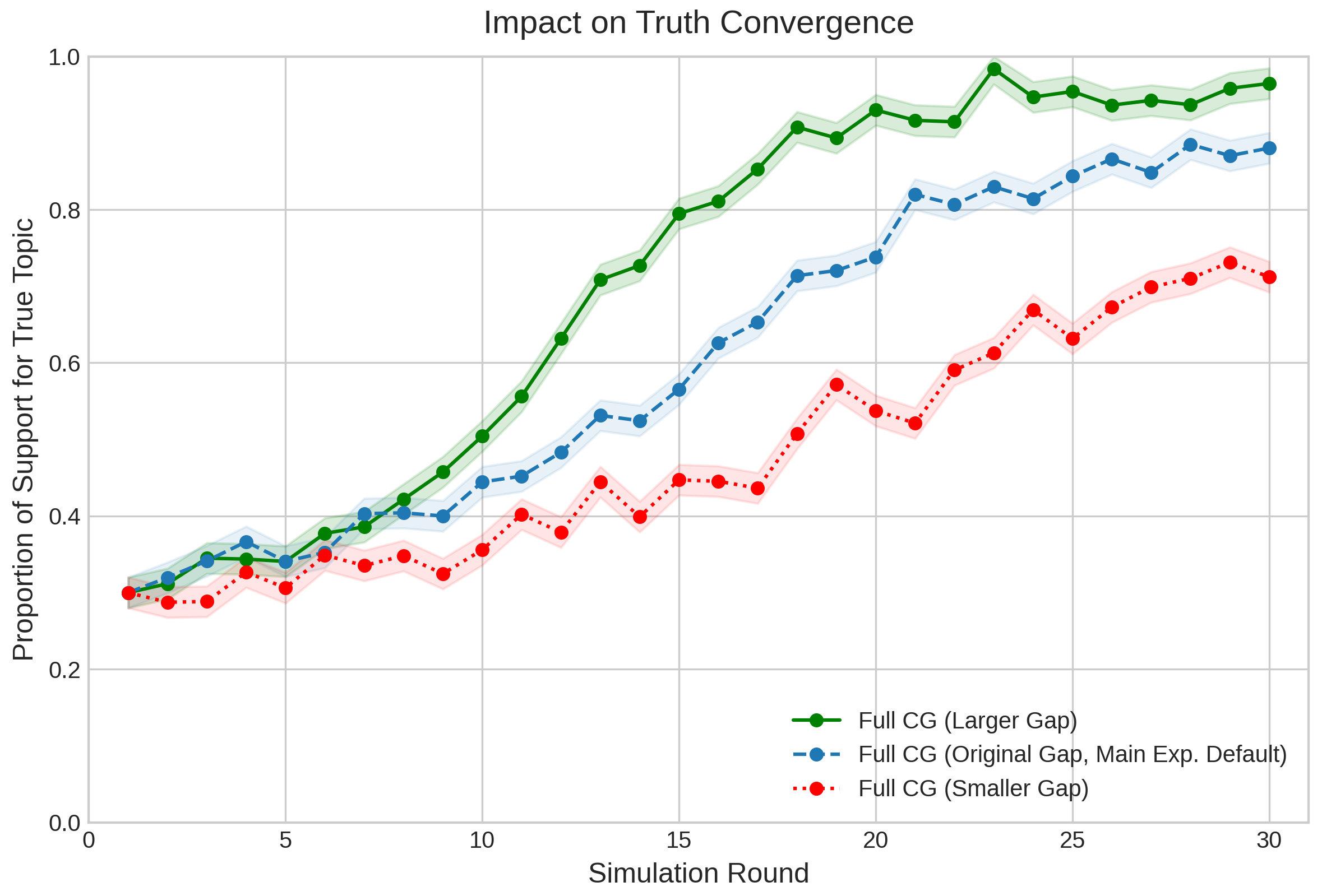}
        \caption{Sensitivity to Baseline Velocity Gap. Comparison of CG's convergence for smaller, original, and larger intrinsic growth differentials.}
        \label{fig:appendix:velocity_gap}
    \end{figure}
    Figure \ref{fig:appendix:velocity_gap} illustrates Credibility Governance's (CG) sensitivity to the intrinsic quality difference between the true and false topics, quantified by the baseline velocity gap (\ensuremath{v_{\text{true}}} - \ensuremath{v_{\text{false}}}). A larger velocity gap leads to significantly faster and more robust convergence. This is because a more pronounced intrinsic advantage for the true topic means its underlying quality is clearer, allowing CG's mechanisms to more effectively detect and amplify this signal, thereby empowering its supporters more efficiently. Conversely, a smaller velocity gap results in slower, more volatile convergence, potentially reaching a lower final truth support. This occurs because a subtle quality difference makes the true topic's advantage harder for CG to discern, as it is more easily masked by noise, the initial resource advantage of the false topic, and individual agent stochasticity. This highlights that while CG is effective, the inherent distinguishability of truth still fundamentally influences its convergence dynamics.

\section{Mapping to Real-World Observable Signals}

Table~\ref{tab:mapping} maps CG variables to concrete, time-indexed proxies in three data-rich examples (forecasting, peer review, and forums). CG only requires (i) a public social signal \(\Theta_k^t\) (and its change \(\Delta\Theta_k^t\)) and (ii) an outcome-linked signal \(\pi_k^t\) that arrives with noise and delay (often observed via a proxy \(\bar{\pi}_k^t\)).

\begin{table*}[t]
  \centering
  \caption{Mapping CG variables to example observable proxies in forecasting, peer review, and community forums.}
  \label{tab:mapping}
  \renewcommand{\arraystretch}{1.12}
  \setlength{\tabcolsep}{6pt}
  \footnotesize
  \begin{tabular}{p{2.35cm} p{4.85cm} p{4.85cm} p{4.85cm}}
  \toprule
  \textbf{CG variable} & \textbf{Forecasting tournaments} & \textbf{Peer review} & \textbf{Community forums} \\
  \midrule
  
  \(\Theta_k^t\) (social signal)
  & Aggregate crowd belief on event \(k\), e.g., mean implied probability or weighted forecast consensus at time \(t\).
  & Aggregate stance toward paper \(k\), e.g., mean recommendation score, accept probability proxy, or discussion sentiment.
  & Aggregate support for claim or topic \(k\), e.g., upvote share, endorsement rate, agreement ratio, or reshare intensity. \\
  
  \(\Delta\Theta_k^t\) (momentum)
  & Change in consensus probability over a window, indicating information arrival or crowd revision.
  & Shift in consensus during deliberation, e.g., score changes after rebuttal or sentiment swing between rounds.
  & Change in endorsement rate or vote share, capturing emerging agreement or coordinated surges. \\
  
  \(\pi_k^t\) (outcome, truth linked)
  & Realized event outcome, resolved binary or continuous outcome (e.g., whether event occurred).
  & Post publication outcomes, e.g., reproducibility checks, later expert assessments, or editor adjudication.
  & Later verification signal, e.g., fact check labels, moderator confirmation, or authoritative source validation. \\
  
  \(\bar{\pi}_k^t\) (blurred outcome proxy)
  & Noisy short horizon proxy such as partial indicators, early reports, or delayed settlement status.
  & Noisy proxies such as reviewer confidence, independent quick checks, or editor triage signals.
  & Noisy proxies such as reports, flags, or limited audits, available with delay. \\
  
  \(w_i^t\) (credibility weight)
  & Forecaster reliability weight used to aggregate predictions, based on past calibration and update dynamics.
  & Reviewer reliability weight used in aggregation or assignment, based on historical consistency and informativeness.
  & User credibility weight used in ranking or visibility, based on past helpfulness and resistance to manipulation. \\
  
  \(r_k^t\) (resource allocation)
  & Attention allocated to event \(k\), e.g., information prominence or forecast visibility weight.
  & Editorial effort on paper \(k\), e.g., additional review bandwidth, expert consultation, or priority handling.
  & Exposure allocated to topic or claim \(k\), e.g., ranking position, recommendation frequency, or distribution budget. \\
  
  \(B_k^t\) (anti bubble penalty)
  & Regularizer against runaway herding, e.g., down weighting sharp swings lacking corroboration.
  & Regularizer against cascade acceptance, e.g., requiring diverse evidence or independent justification before promotion.
  & Regularizer against viral cascades, e.g., reducing amplification when engagement growth exceeds evidence growth. \\
  
  Noise and delay
  & Outcomes resolve with delay, and interim signals can be noisy or adversarially influenced.
  & Ground truth is delayed, and reviewer signals are noisy, biased, and socially influenced.
  & Verification is delayed, and engagement signals can be manipulated by coordinated actors. \\
  
  \bottomrule
  \end{tabular}
  \end{table*}

    \section{Robustness Testing}

\subsection{H4: CG is most beneficial at intermediate noise, by reducing lock-in and speeding correction}
\label{sec:h4_noise}

\begin{figure}[h]
  \centering
  \includegraphics[width=\linewidth]{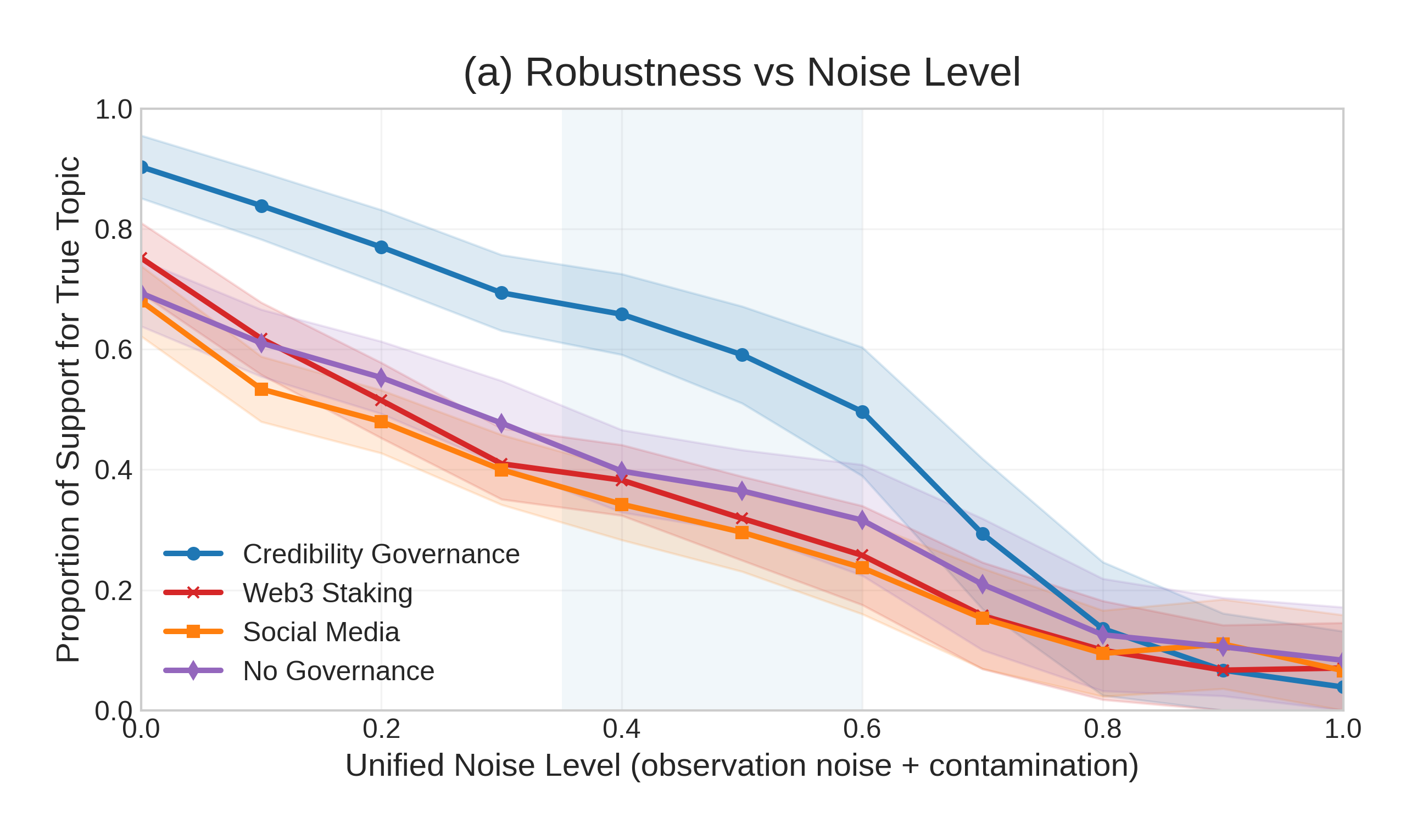}
  \caption{\textbf{H4, robustness vs noise.} We sweep a unified noise level combining observation noise and contamination. CG shows the largest accuracy gains in the intermediate-noise regime, where truth is weakly observable yet recoverable. Under extreme noise, all mechanisms degrade.}
  \label{fig:noise_sweep_main}
\end{figure}

\begin{figure}[h]
  \centering
  \includegraphics[width=0.86\linewidth]{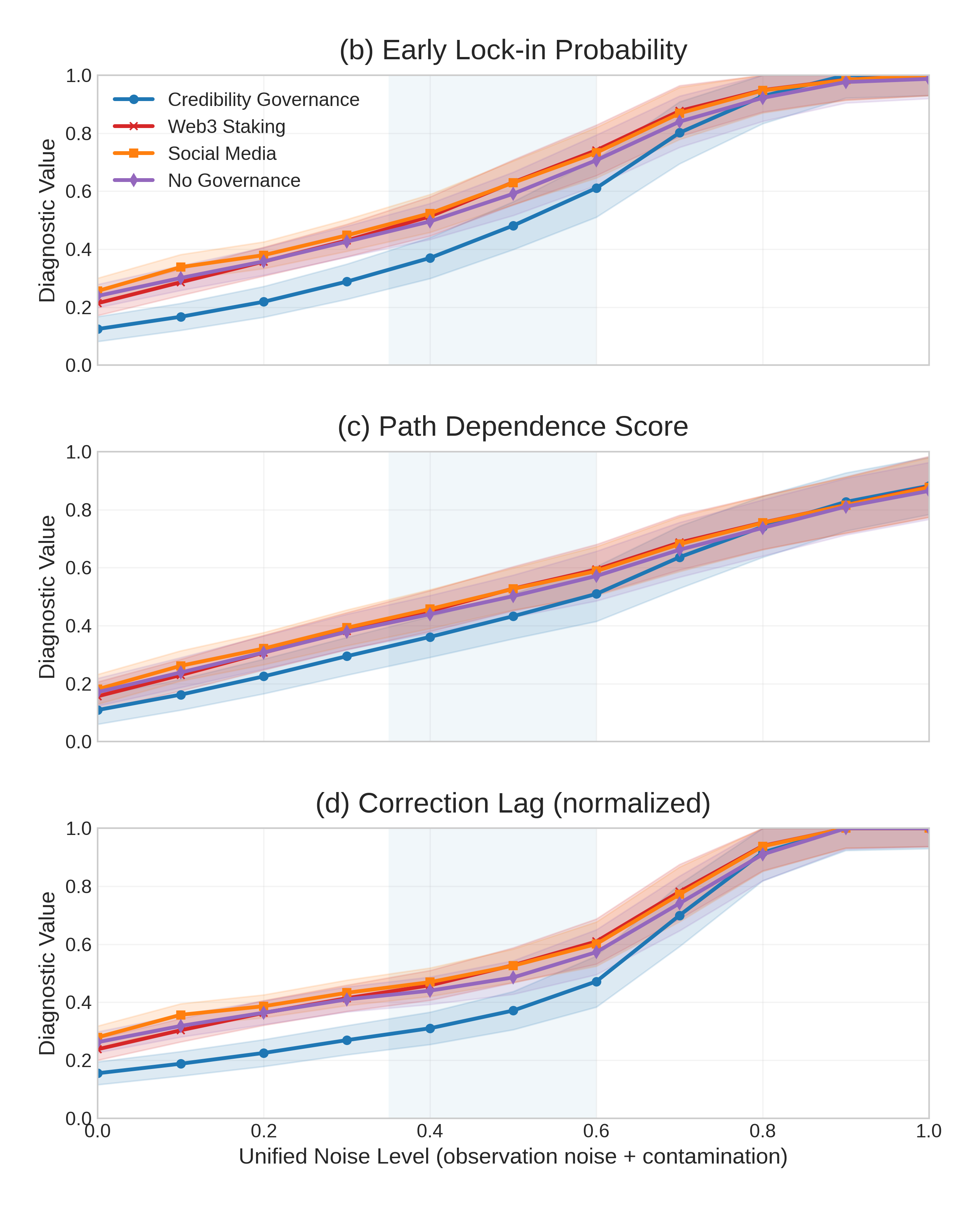}
  \caption{\textbf{H4, diagnostics under noise.} As noise increases, we track early lock-in probability, path dependence score, and correction lag (normalized). CG reduces lock-in and path dependence and shortens correction lag most strongly in the intermediate-noise regime.}
  \label{fig:noise_diagnostics}
\end{figure}

\noindent
\textbf{Hypothesis (H4).} CG's advantage peaks at intermediate noise, where naive popularity mechanisms lock in early mistakes, but evidence remains informative enough for correction. \\
\textbf{Test.} We sweep a unified noise level and measure accuracy (support for the true topic), plus diagnostics for early lock-in, path dependence, and correction lag. \\
\textbf{Result.} Fig.~\ref{fig:noise_sweep_main} shows CG achieves the largest improvement under intermediate noise, while all methods degrade at extreme noise. Fig.~\ref{fig:noise_diagnostics} shows that in the same noise band, CG lowers early lock-in and path dependence and shortens correction lag, indicating that CG improves accuracy by suppressing cascade-driven failure modes and enabling faster correction.

\subsection{H5: Explicit credibility prompting provides a modest additional benefit}
\label{sec:h5_prompting}

\begin{figure}[h]
  \centering
  \includegraphics[width=0.86\linewidth]{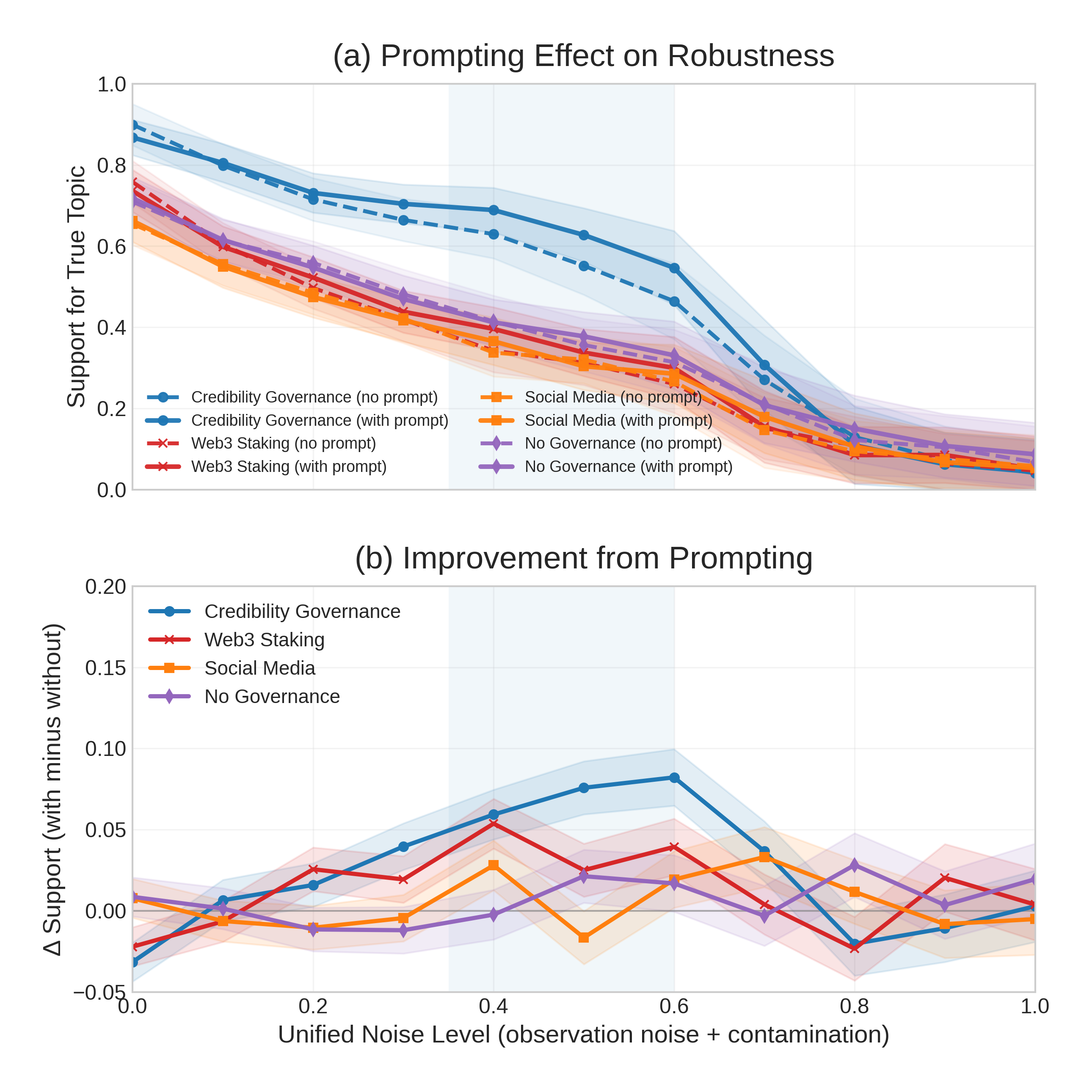}
  \caption{\textbf{H5, prompting effect.} Solid lines denote runs with explicit credibility prompting and dashed lines denote runs without. Prompting yields a modest, consistent uplift, with the largest improvement under intermediate noise.}
  \label{fig:prompting_effect}
\end{figure}

\noindent
\textbf{Hypothesis (H5).} Explicitly prompting agents to consider credibility offers incremental gains on top of incentive design, especially when signals are ambiguous. \\
\textbf{Test.} We compare outcomes with versus without explicit credibility prompting across the noise sweep. \\
\textbf{Result.} Fig.~\ref{fig:prompting_effect} shows a modest but consistent improvement with prompting, peaking under intermediate noise. This suggests CG's incentives already induce effective credit assignment, while prompting further stabilizes reasoning under ambiguity.

\begin{figure*}[h]
  \centering
  \includegraphics[width=0.8\linewidth]{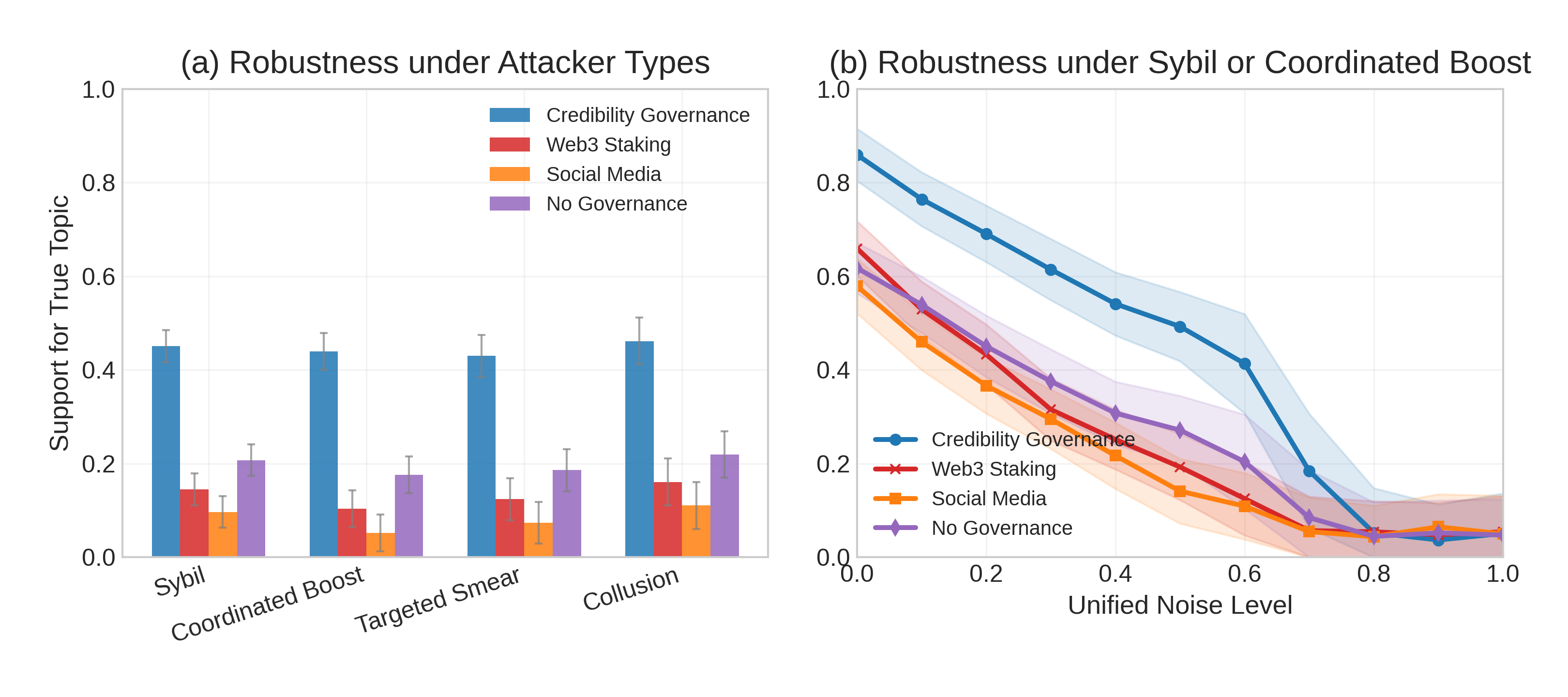}
  \caption{\textbf{H6, threat model.} (a) Robustness under multiple attacker behaviors at a representative noise level. (b) Robustness versus noise under representative attack types. CG degrades less across attacker types than WS and SM, which are more sensitive to manipulation that creates artificial momentum.}
  \label{fig:threat_model}
\end{figure*}

\begin{figure*}
  \centering
  \includegraphics[width=0.8\linewidth]{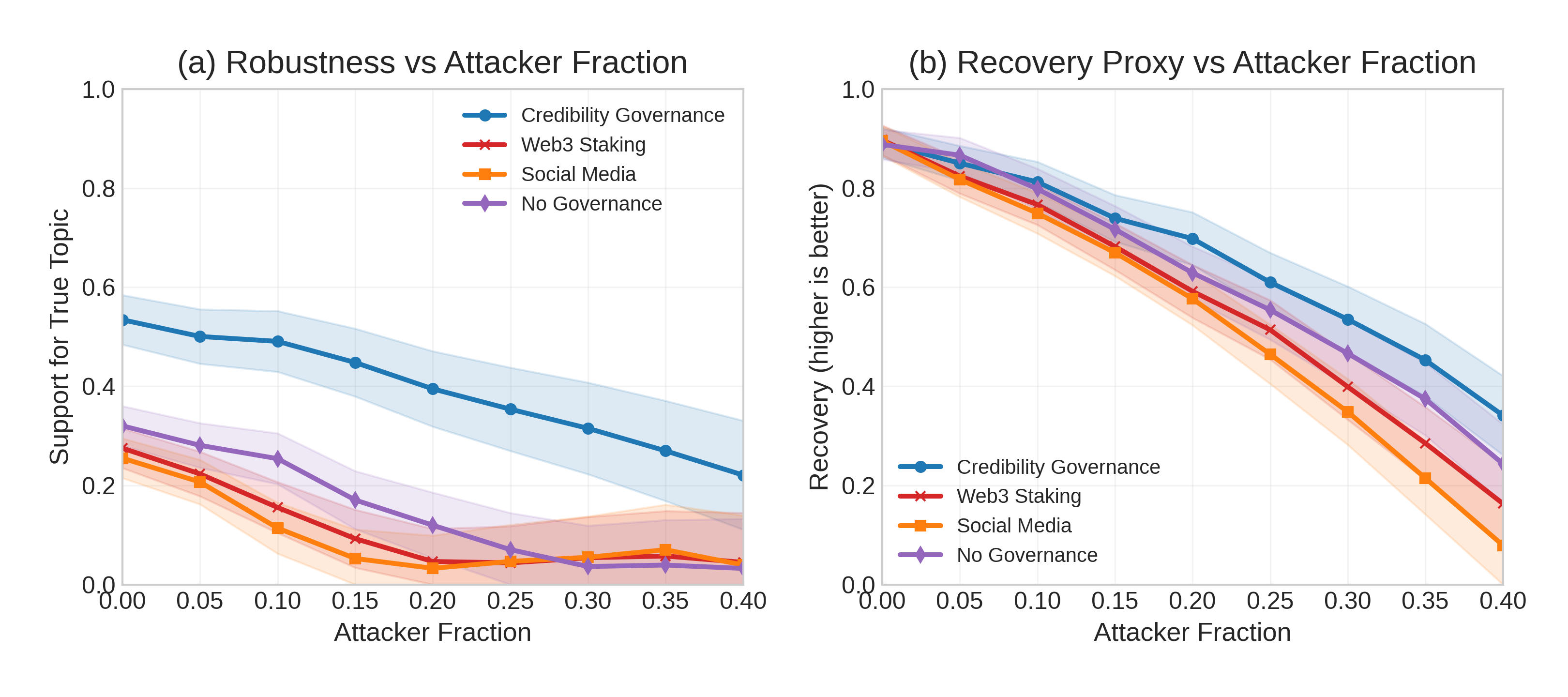}
  \caption{\textbf{H6, attacker fraction sweep.} (a) Accuracy versus attacker fraction at a representative noise level. (b) A recovery proxy versus attacker fraction. CG tolerates larger adversarial populations before collapsing and maintains better recovery under higher attacker prevalence.}
  \label{fig:attacker_fraction}
\end{figure*}

\begin{figure*}
  \centering
  \includegraphics[width=0.8\linewidth]{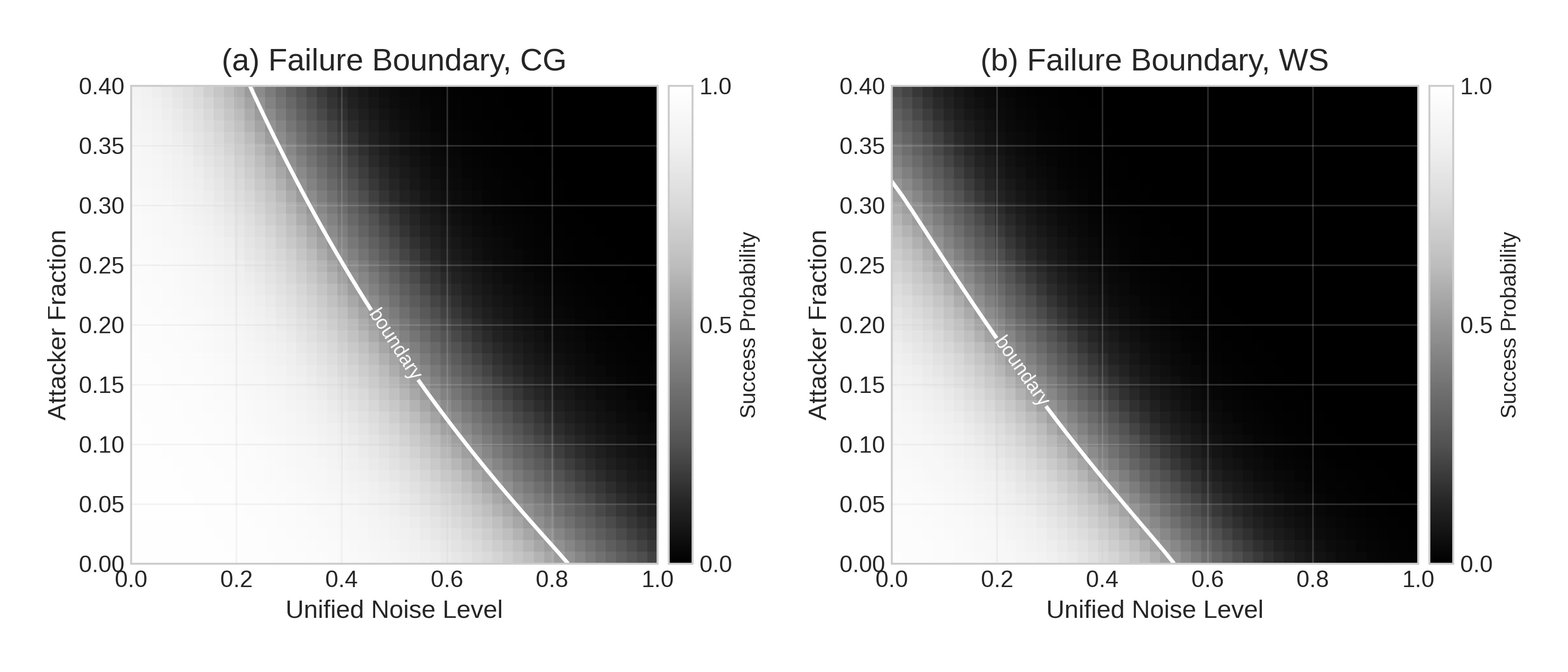}
  \caption{\textbf{H6, failure boundary over noise and attacker fraction.} Heatmaps show success probability as a function of unified noise and attacker fraction, with the white contour indicating the boundary (success probability $=0.5$). CG shifts the boundary toward higher noise and higher attacker fractions compared to WS, expanding the stable operating region.}
  \label{fig:failure_boundary}
\end{figure*}

\subsection{H6: CG expands the safe operating region under adversarial manipulation}
\label{sec:h6_adversary}

\noindent
\textbf{Hypothesis (H6).} CG is more robust to strategic manipulation than momentum-amplifying baselines, and expands the operating region where collective learning remains stable under joint noise and adversarial pressure. \\
\textbf{Test.} We evaluate (i) multiple attacker behaviors (threat model), (ii) attacker prevalence sweeps, and (iii) a 2D failure boundary over noise and attacker fraction. \\
\textbf{Result.} Fig.~\ref{fig:threat_model} shows CG degrades less across attacker types. Fig.~\ref{fig:attacker_fraction} shows CG tolerates larger adversarial fractions before accuracy collapses and maintains better recovery. Fig.~\ref{fig:failure_boundary} summarizes the joint boundary and shows CG has a larger stable operating region than WS, shifting the boundary toward higher noise and higher attacker fractions.


\noindent
\textbf{Summary.} Across hypotheses, CG improves collective accuracy by reallocating influence toward truth-aligned agents using momentum in public evidence, suppressing early lock-in and path dependence, and expanding robustness under both noise and adversarial manipulation.

\end{document}